\documentclass[twocolumn,iop,revtex4]{openjournal}

\usepackage{newtxtext,newtxmath}

\usepackage[T1]{fontenc}

\DeclareRobustCommand{\VAN}[3]{#2}
\let\VANthebibliography\thebibliography
\def\thebibliography{\DeclareRobustCommand{\VAN}[3]{##3}\VANthebibliography}

\usepackage{graphicx}	
\usepackage{amsmath}	
\usepackage{soul}
\usepackage{multirow}
\usepackage{natbib} 
\usepackage{hyperref}
\usepackage{amssymb}
\usepackage[title]{appendix}
\hypersetup{colorlinks=true,linkcolor=blue,citecolor=blue,filecolor=blue,urlcolor=blue}

\newcommand{\Msun}{M_{\odot}}

\begin{document}

\title[Halo Masses]{Halo Properties from Observable Measures of Environment: II. Central versus Satellite Classification}

\author{
Haley Bowden,$^{1}$\thanks{E-mail: hbowden@arizona.edu}
Peter Behroozi,$^{1}$}

\affiliation{
$^{1}$Department of Astronomy and Steward Observatory, University of Arizona, Tucson, AZ 85721, USA
}

\begin{abstract}

A physical understanding of galaxy formation and evolution benefits from an understanding of the connections between galaxies, their host dark matter halos, and their environments. In particular, interactions with more-massive neighbors can leave lasting imprints on both galaxies and their hosts. Distinguishing between populations of galaxies with differing environments and interaction histories is therefore essential for isolating the role of environment in shaping galaxy properties. We present a novel neural-network based method, which takes advantage of observable measures of a galaxy and its environment to recover whether it (1) is a central or a satellite, (2) has experienced an interaction with a more massive neighbor, and (3) is currently orbiting or infalling onto such a neighbor.  We find that projected distances to, redshift separations of, and relative stellar masses with respect to a galaxy’s 25 nearest neighbors are sufficient to distinguish central from satellite halos in $> 90\%$ of cases, with projection effects accounting for most classification errors. Our method also achieves high accuracy in recovering interaction history and orbital status, though the network struggles to distinguish between splashback and infalling systems in some cases due to the lack of velocity information. With careful treatment of the uncertainties introduced by projection and other observational limitations, this method offers a new avenue for studying the role of environment in galaxy formation and evolution.

\end{abstract}




\section{Introduction}\label{sec:Intro}

In the $\Lambda$CDM cosmological paradigm of structure formation, galaxies form within dark matter halos. These halos are gravitationally self-bound structures that grow hierarchically through mergers of smaller halos into increasingly massive structures, with the largest largest halo in a given overdensity (i.e., the central halo) potentially containing many smaller self-bound structures (i.e., satellite halos). The properties of galaxies residing within these halos, such as their sizes and star formation rates, are strongly influenced by the assembly history and environmental context of their host halos (see \citealt{Wechsler_2018} for a review). Understanding the galaxy-halo connection is crucial to using galaxy properties as tracers of dark matter properties and histories.

To explore the influence of environment and the galaxy-halo connection on galaxy formation and evolution, it is often necessary to distinguish central galaxies from those residing in satellite halos, as the two populations are subject to different physical processes. Central galaxies may continue to grow through accretion, and, when they do stop forming stars, are thought to quench primarily through primarily internal processes. In contrast, satellites fall within the sphere of influence of a more massive halo and are thereby subject to tidal forces from the central, as well as interactions with other satellites \citep{Peng_2012,Bluck_2020}. These factors lead to satellite-specific quenching mechanisms include processes such as ram pressure stripping, tidal stripping, and galaxy-galaxy harassment, which have a strong dependence on environment \citep{Peng_2012}.  

This distinction is reflected prominently in the different observed quenched fractions of central and satellite galaxies at the same stellar mass (\citealt{Bosch_2008,Wetzel_2012,Woo_2015,Bluck_2016}). Additionally, its roots in environmental versus internal quenching dominance is supported by the strong observed correlations between satellite quenching and local environmental density, whereas quenching in central galaxies is only weakly associated with environment (\citealt{Kakos_2024}). However, alternative interpretations propose that observed differences between centrals and satellites primarily reflect variations in the underlying stellar-to-halo mass relations rather than distinct quenching mechanisms (\citealt{Wang_2018a, Wang_2018b, Wang_2020_CvS}).

A key limitation in identifying the physical mechanisms responsible for differences between central and satellite galaxies arises from challenges in observationally distinguishing these two populations. Central galaxies have often been selected based on an isolation criteria (e.g., \citealt{More_2011,Duplancic_2018,Mesa2021,Gu_2024}), where the brightest galaxy in a region is identified as a central if it has no neighbors of similar or greater magnitude within a cylindrical aperture. Satellites are then similarly identified as fainter galaxies within the same aperture. As discussed in \cite{Gu_2024}, for a cylindrical aperture based method, misidentification can occur due to the projection of foreground/background galaxies into the aperture, as well as cases where a satellite may be brighter than its central. \cite{Campbell_2015} suggests that the latter could occur in $\sim$10\% of galaxy groups at masses of $10^{13}\Msun$, with increasing frequency at higher masses.

An alternative approach, particularly for larger systems, is to use a group catalog (e.g., \citealt{Yang_2007,Tinker_2011}), where the most luminous galaxy in the group is then assigned as the central. As with the isolation criteria, this method is subject to misidentifications when the central is not the brightest galaxy in the group. Additionally, errors in the group catalog can lead to fracturing or merging of groups. \citet{Campbell_2015} explores how these factors result in the misidentification of centrals and satellites, thus biasing the observed trends in quenching for the two populations.

In this paper, we present a new method for distinguishing between central and satellite galaxies in observations, aiming to reduce misclassification. Our approach leverages information about the stellar masses and relative positions of neighboring galaxies, capitalizing on the fact that halo properties and formation histories are tied to their local environment (see, e.g., \citealt{Behroozi_2021}). We employ a neural network to extract this information from a high-dimensional set of galaxy positions and stellar masses, refining the classification process.

While most satellite definitions use some form of spherical-overdensity criterion, there remains a question of what radius to use and whether a spherical overdensity definition fully captures the physics of the central/satellite system. For example, the sphere of influence of a halo is thought to extend far past its virial radius, with neighboring galaxies at larger separations experiencing mass loss and preferential quenching (\citealt{Bahe_2013, Behroozi_2014, Fong_2021, Lacerna_2022}). The splashback radius, defined as the apocenter of particles on their first orbit, has been proposed as a more physical boundary than the virial radius (\citealt{Balogh_2000, Bahe_2013, Wetzel_2014, Diemer_2021}). Between the virial radius and the larger splashback radius exists a large population of ‘backsplash’ halos, which formerly fell within the virial radius of a more-massive host, but continued on an orbital trajectory that brought them outside that host’s virial radius. These halos and their associated galaxies tend to retain a significant signature of their interaction with their former host, marking them as distinct from genuine isolated centrals (\citealt{Knebe_2011, Muriel_2014, Diemer_2021, Borrow_2023}).

Using a larger radius, such as the the splashback radius, for the spherical overdensity criterion, naturally includes backsplash halos as satellites, but also indiscriminately includes a population of infalling halos and genuine flyby events that have not, as of the current snapshot, experienced a significant interaction with a more-massive halo. Examination of the phase-space of the halo-subhalo population demonstrates that no spherical-overdensity criteria alone can fully separate the population of backsplash subhalos from those on first-infall (\citealt{Diemer_2021, Diemer_2022, Garcia_2023}). Taking this into account, we also explore a classification scheme in this paper that includes backsplash halos as satellites by checking whether objects were considered satellites in previous simulation snapshots, while maintaining the strict $R_{vir}$ spherical overdensity criteria to reduce contamination from infalling halos and true flybys. 

The issue of defining a halo boundary has also been explored analogously in separating infalling and orbiting particles in simulations. \cite{Diemer_2022} considers a split based on pericentric passage, while \cite{Garcia_2023} splits particles based on accretion time, both of which are in general agreement. In this paper, we consider an analogous approach to \cite{Garcia_2023}, classifying subhalos into orbiting and infalling populations based primarily on accretion time.  We expect some of this information to be recoverable from observations of the stellar mass and positions/redshifts of neighboring galaxies, however, limitations in the observations, particularly regarding the relative 3D velocity between objects, do present challenges in constraining the trajectories and histories of individual halos.

This work focuses on the application of environmental information to the problem of central and satellite classification and is organized as follows. Section \ref{sec:Data} provides an overview of the mock halo and galaxy catalogs used throughout. In Section \ref{sec:Def}, we discuss different criteria for splitting our halos into two populations, including current centrals and satellites. Section \ref{sec:Methods} contains an overview of different measures of the local environment our methodology for halo classification, while Section \ref{sec:Results} covers the results of these methods. Lastly, Sections \ref{sec:Discussion} and \ref{sec:Conclusions} contains a discussion of our results and future directions for this work. We adopt a standard $\Lambda$CDM cosmology with $(h, \Omega_m, \sigma_8, n_s) = (0.678, 0.307, 0.823, 0.96)$ throughout.

\section{Data}\label{sec:Data}

Our method for classifying halos leverages data from simulations. For training and evaluation the `true' classification of a halo is determined based on information in the halo merger tree. Training and optimization are performed on the \textit{Small MultiDark Planck (SMDPL)} simulation, with performance testing on the \textit{Bolshoi-Planck} cosmological simulation (Section \ref{subsec:HaloProps}; \citealt{Bolshoi,Rodriguez_Puebla_2016}). Our classification methods rely on galaxy stellar masses and positions, which correspond closely to observable quantities in galaxy surveys. Galaxy stellar masses are assigned according to the  \textsc{UniverseMachine} empirical model (Section \ref{subsec:GalProps}; \citealt{UM}).

\subsection{Dark Matter Simulations}\label{subsec:HaloProps}
Optimization and network training were performed on the $z=0$ snapshot from the SMDPL simulation \citep{SMDPL}. This simulation box has a periodic volume of (400 $h^{-1}$Mpc)$^3$ and 3840$^3$ particles, achieving a mass resolution of $9.63 \times 10^7 h^{-1}M_{\odot}$ per particle and a force resolution of 1.5 $h^{-1}$kpc. The SMDPL simulation adheres to a flat $\Lambda$CDM model consistent with the latest observations from Planck, with $(h, \Omega_m, \sigma_8, n_s) = (0.678, 0.307, 0.823, 0.96)$.  We assume the same cosmology throughout this work.

Halo finding was carried out using \textsc{Rockstar} \citep{Rockstar}. The \textsc{Rockstar} halo finder classifies halos as centrals or satellites (i.e., subhalos) based on whether the halo falls within the virial radius of a more-massive halo. We retained this definition for our classification. Merger histories were constructed for the identified halos via the \textsc{ConsistentTrees} code \citep{ConsistentTrees}. Halo masses and virial radii were defined based on the \citet{Bryan1998} virial spherical overdensity definition.

To validate our models, we test them on the $z=0$ snapshot from the \textit{Bolshoi-Planck} simulation \citep{Bolshoi}. This is a smaller box than SMDPL with a co-moving volume of (250 $h^{-1}$Mpc)$^3$ and 2048$^3$ particles, yielding a mass resolution of $1.55 \times 10^8 h^{-1}M_{\odot}$ per particle and a force resolution of 1.0 $h^{-1}$kpc. This simulation employs a cosmology comparable to that of the SMDPL, with  $(h, \Omega_m, \sigma_8, n_s) = (0.68, 0.30711, 0.82, 0.96)$, ensuring consistency across our datasets.

\subsection{Mock Galaxy Catalogs}\label{subsec:GalProps}
The properties of the individual galaxies which populate the halos were derived from \textsc{UniverseMachine} \citep{UM}, an empirical model describing the galaxy-halo connection. \textsc{UniverseMachine} uses a Markov Chain Monte Carlo algorithm to constrain the evolution of galaxy star formation rates over cosmic time as a function of halo mass and halo growth rates. This approach ensures the simulated galaxy population aligns with observed data across various metrics, including the stellar mass function, cosmic star formation rates, and specific star formation rates, among others (as detailed in Appendix C of \citealt{UM}).

We extracted galaxy positions, velocities, and stellar masses from the \textsc{UniverseMachine} mock galaxy catalogs corresponding to the selected simulation snapshots. We select the so-called ‘observed’ stellar masses from the catalog. These mass values have been adjusted from the true stellar masses to account for systematic discrepancies and observational variance \citep{UM}. 

Although this work exclusively uses \textsc{UniverseMachine} stellar masses, the methodology is designed for broader application to observationally derived stellar masses. In practice, measurements of stellar mass are model and calibration dependent \citep[e.g.,][]{Conroy2013,Madau2014,Mobasher2015}. To mitigate some of the systemic biases inherent in different stellar mass estimation models, we convert stellar masses to cumulative number densities (i.e., relative mass rankings). This is accomplished by ordering the galaxies within a given simulation box by stellar mass, with the most massive being assigned a rank of one. These rankings are then normalized by the box volume. This normalization process helps ensure the applicability of our neural network model across a broader range of datasets.

\section{Defining Halo Populations}\label{sec:Def}

While it is common practice to separate galaxies and their host halos into satellite and central populations, the dividing line can be drawn in many different ways (e.g., spatially or dynamically), each being better suited to certain kinds of analysis. Additionally, the question remains of whether a central/satellite split is always the most valuable way to classify these halo populations. Within this paper, we consider three different categorizations as outlined below.

An essential property of satellites is that they are subject to tidal forces from the central potential leading to tidal disruption and mass stripping. The tidal force exerted on an object by its neighbors is a monotonic function of $R/R_\text{vir, neighbor}$ (i.e., the deeper the object plunges into the potential well of the larger halo, the stronger the tidal force). Thus, the magnitude of the tidal force experienced by an object can serve as a continuous measure for how closely the object resembles a satellite, including objects outside the virial radius. With this in mind, it is useful to define the \textit{halo-centric distance} (HCD) to each neighbor as the ratio of the 3D separation between the objects to the neighbor's virial radius ($R/R_\text{vir, neighbor}$). The minimum value of the halo-centric distance across an object’s neighbors corresponds to the normalized distance to the neighbor exerting the largest tidal force on the object. 

\subsection{Centrals vs. Satellites (Present)}\label{subsec:Def1}
\noindent{}\textbf{A halo is considered a satellite if it falls within the virial radius of a more-massive halo at the current snapshot. Otherwise, it is a central.}

\begin{figure}
     \centering
     \includegraphics[width=0.45\textwidth]{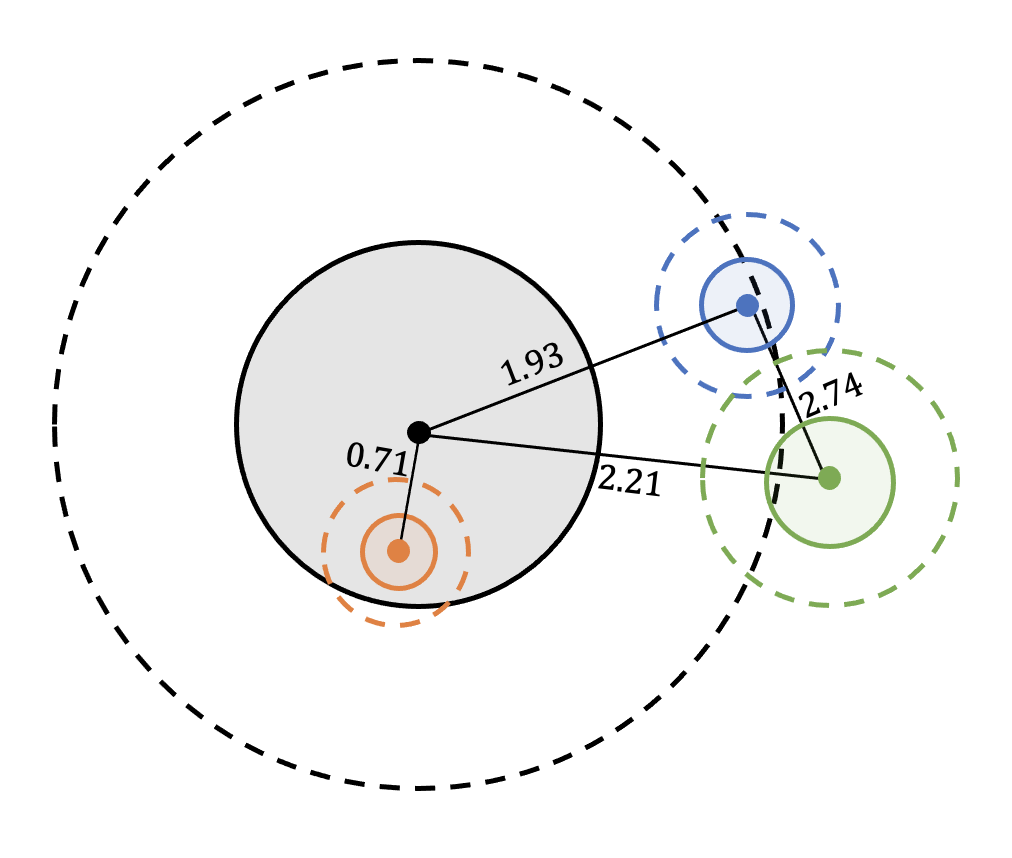}
     \caption{Here we show a 2D projection of four objects, each centered at a point with its virial radius indicated by a shaded circle surrounding that point and a dashed line representing 2R$_{\text{vir}}$ enclosing both. The values shown along the connecting lines between objects represent the halo-centric distance, i.e., the distance between the two halo scaled by the virial radius of the more-massive halo. By this definition, the orange halo clearly falls within 1 $R_{vir}$ of the black halo (it has a halo-centric distance $< 1$ with respect to the black halo), with no other nearby-larger objects, and thus is a satellite of the black halo. On the other hand, the blue and green halos do not meet this criteria. Additionally, the blue halo, while physically closer to the green halo, has a smaller halo-centric distance relative to the black halo, due to its much larger size, making the black halo the object exerting the largest tidal force on the blue halo.
     }
     \label{fig:HCD}
\end{figure}

The first classification scheme follows a standard description of the substructure of a halo, with the extent of the host halo (central) defined by its virial radius and subhalos (satellites) as smaller bound structures located within that radius. For each object within our simulation box, we search for nearby more massive halos. If the halo-centric distance to a neighbor is less than one, the object must necessarily fall within the virial radius of that more-massive neighbor, and thus is a satellite at the current snapshot. Figure \ref{fig:HCD} illustrates how the minimum halo-centric distance is assigned for each halo, and how this value can be used to distinguish between satellites (orange object) and centrals (other objects).

\begin{figure}
     \centering
     \includegraphics[width=0.45\textwidth]{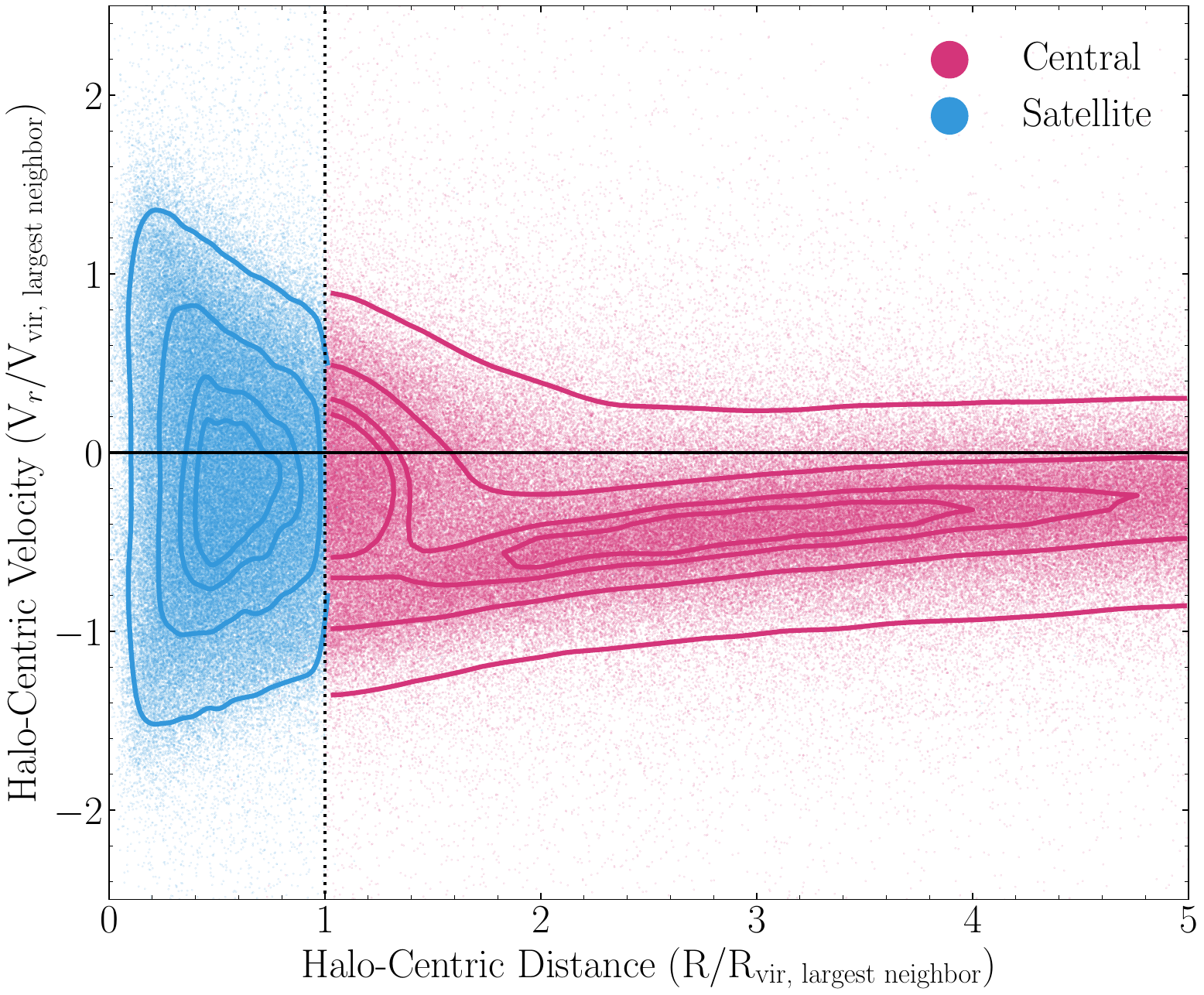}
     \caption{Halos from the Bolshoi-Planck simulation are plotted according to their halo-centric distance (x-axis) and relative radial velocity (y-axis) to the neighbor exerting the largest tidal influence. Satellites are represented by blue points and centrals by pink. The dividing line between the two classes is the dotted vertical line, which is set by the point when the distance between a halo and its more-massive neighbor is less than virial radius of said neighbor. Moving outwards, contours contain 15\%, 30\%, 60\%, and 90\% of the population.
     }
     \label{fig:def1}
\end{figure}

Figure \ref{fig:def1} shows the location of the central and satellite populations on a phase-space diagram for the Bolshoi-Plank simulation box. All satellites (blue) have minimum halo-centric distances of less than one and thus are found to the left of the vertical dashed line in the diagram. This is independent of the relative radial velocities of the object and its neighbor. Within the Bolshoi-Plank and SMDPL simulation boxes, $\sim 25\%$ of galaxies with $M_{*}>10^9 M_{\odot}$ have host-halos classified as current satellites under this scheme.

This classification tells us whether a given galaxy (and its host halo) are \textit{currently} strongly gravitationally influenced by a more massive halo. This is especially relevant when discussing short timescale processes, such as ram-pressure stripping, and the future evolution of the object. Yet, this position-based approach does not fully capture the complexity of the history or dynamics of all subhalos. For example, the halo may be passing through but not significantly disturbed by the `host' halo or it may historically have been dynamically influenced by the `host' halo but not currently fall within the virial radius of that halo. These situations are important to consider when comparing star formation and quenching in central and satellite galaxies. These caveats lead us to our alternate classification schemes.

\subsection{Centrals vs. Satellites (Historical)}\label{subsec:Def2}
\noindent{}\textbf{A halo is considered a satellite if it fell within the virial radius of a more massive halo at any snapshot up to and including the current snapshot. Only halos which have never been within the virial radius of a more massive halo across all snapshots are considered centrals.}

\begin{figure}
     \centering
     \includegraphics[width=0.45\textwidth]{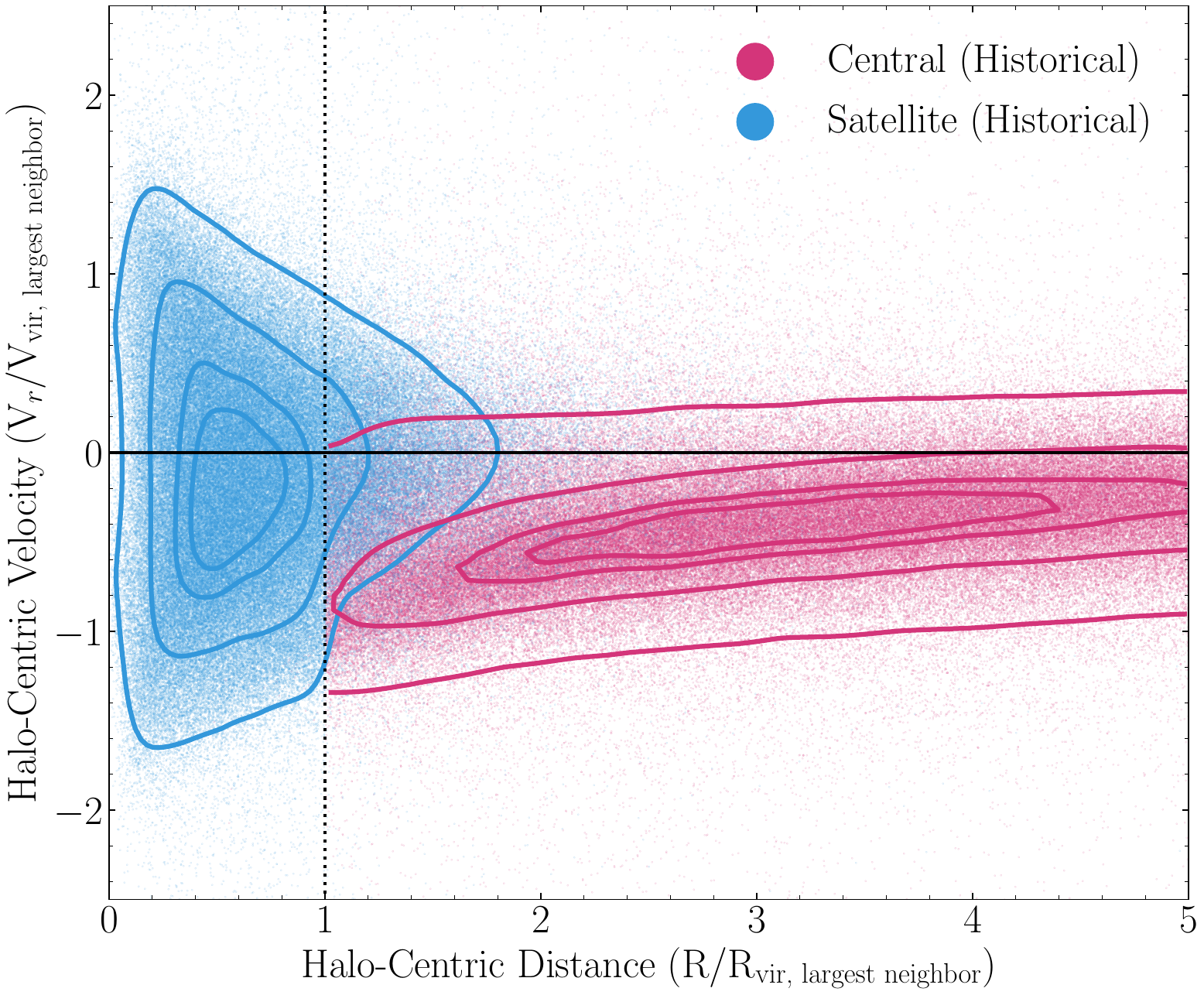}
     \caption{Halos from the Bolshoi-Planck simulation are plotted according to their halo-centric distance (x-axis) and relative radial velocity (y-axis) to the neighbor exerting the largest tidal influence. Historical satellites are represented by blue points and centrals by pink. Moving outwards, contours contain 15\%, 30\%, 60\%, and 90\% of the population. In contrast to Figure \ref{fig:def1}, historical satellites can be found for R/R$_{\text{vir}}$ > 1, i.e., to the right of dotted vertical line. However, all halos within the radial cut remain classified as satellites.
     }
     \label{fig:def2}
\end{figure}

Figure \ref{fig:def2} shows the relative positions and velocities of each object to the neighboring halo exerting the largest tidal force on it at the current snapshot, colored by historical central and satellite status. As in the classification based solely on the current snapshot, all objects to the left of the vertical dashed line (HCD $\le 1$) are satellites. However, additional objects to the right of the line are now also considered historical satellites if they fell to the left of the line in a previous snapshot. This classification adds an additional $\sim 7\%$ of objects to the satellites category compared to the current satellites classification for a total of $\sim 32\%$ satellites (historical) in the simulation box at $z=0$.

As this classification includes post-pericenter objects outside the virial radius of their host (i.e., backsplash halos) as satellites, it is an advantageous scheme if one wishes to consider a larger range of interaction timescales. This scheme is also helpful if one is looking to create a `pure' sample of central halos, which have been free from significant interactions with a larger halo during their evolution.

\subsection{Infalling vs. Orbiting}\label{subsec:Def3}
\noindent{}\textbf{A subhalo is `orbiting' if it was accreted by a more-massive halo more than half a dynamical time ago and satisfies additional velocity and position criteria. Otherwise it is considered `infalling.'}

\begin{figure}
     \centering
     \includegraphics[width=0.45\textwidth]{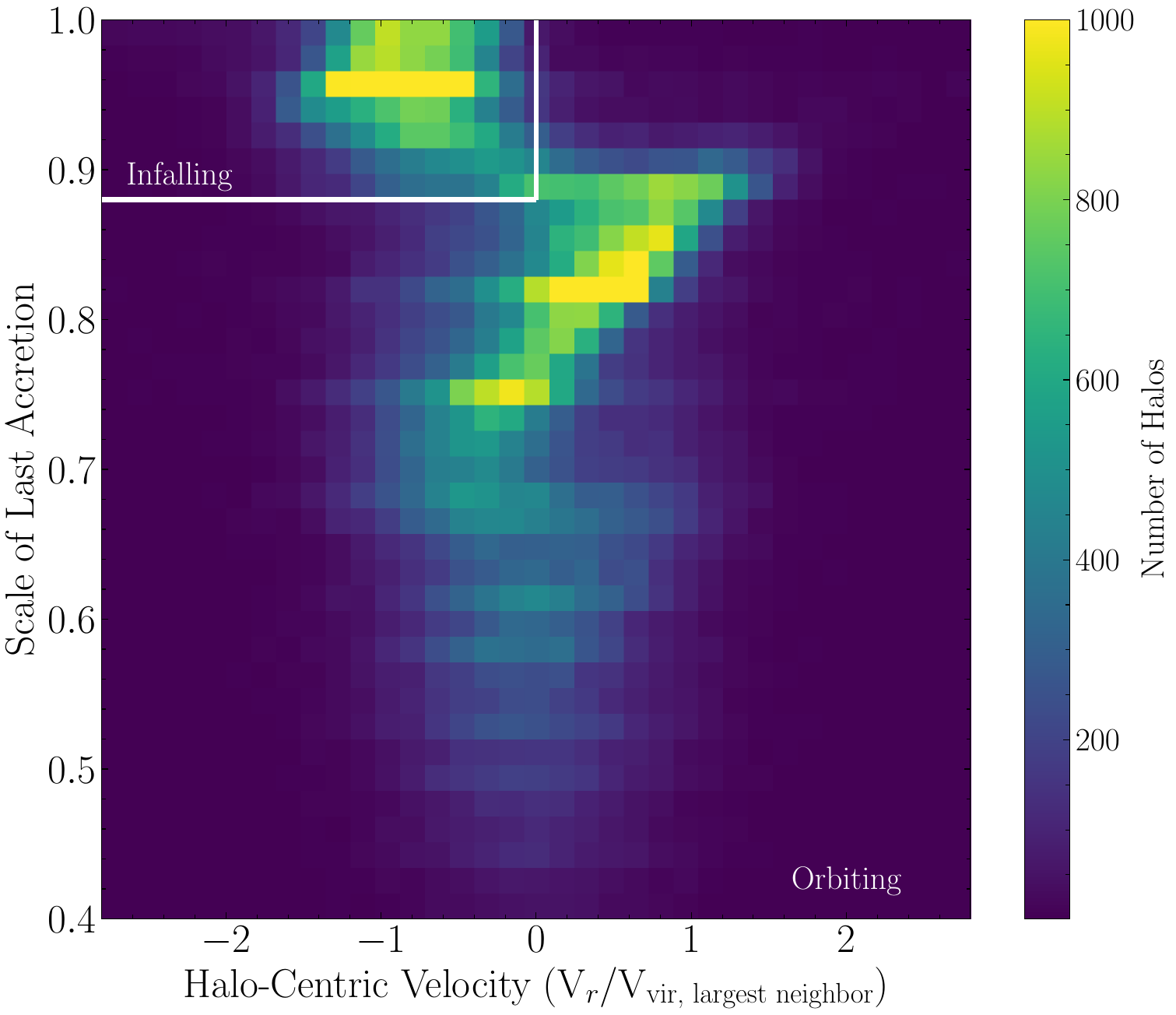}
     \caption{The number density of halos with a halo-centric distance of less than one is shown in relative velocity and accretion-time space. Only halos falling in the top-left box, which have a recent accretion scale ($a_\text{acc} > 0.87$) and a negative radial velocity with respect to their future host are considered infalling. All other objects with halo-centric distance less than one are considered orbiting.
     }
     \label{fig:def3b}
\end{figure}

\begin{figure}
     \centering
     \includegraphics[width=0.45\textwidth]{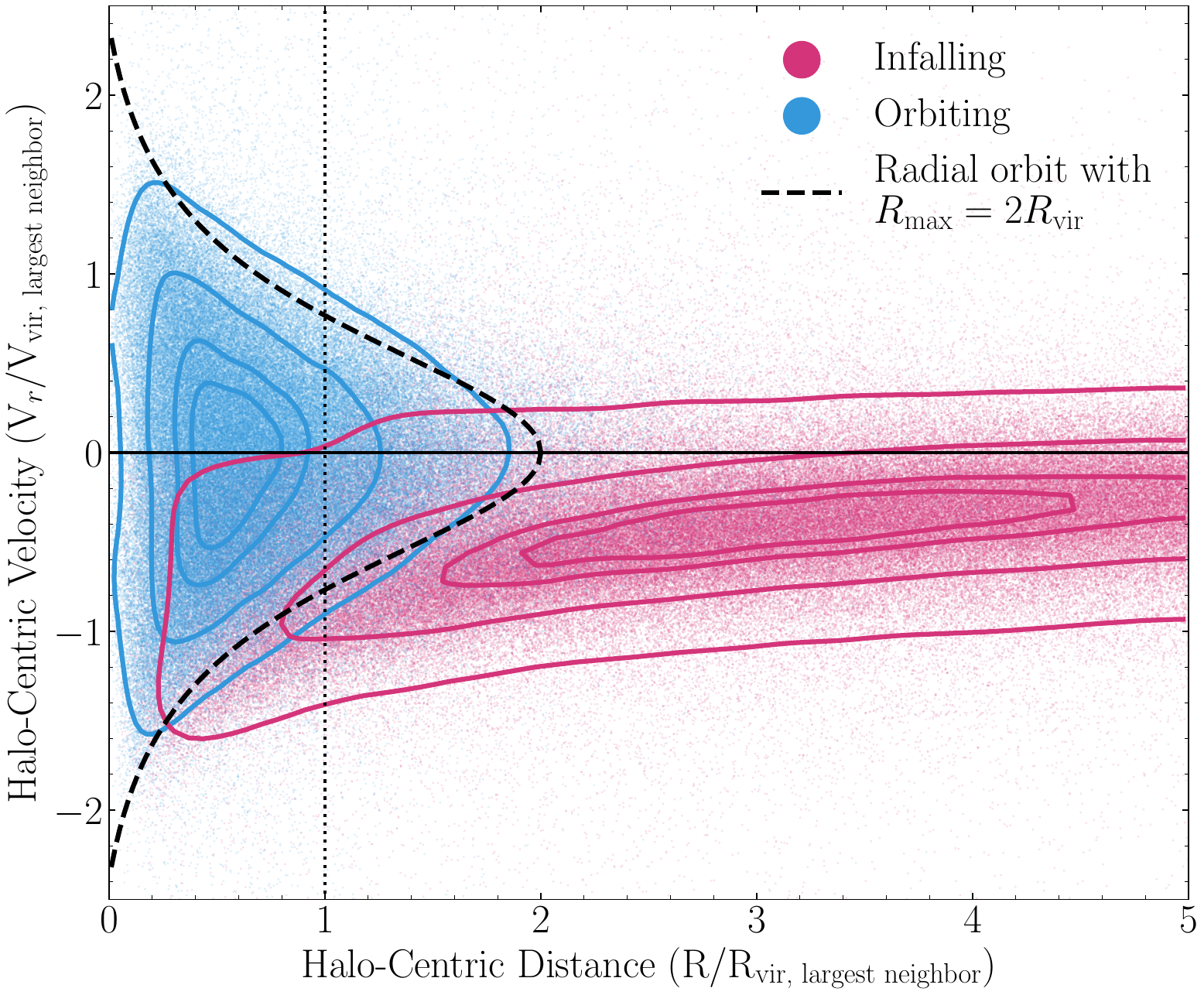}
     \caption{Halos from the Bolshoi-Planck simulation are plotted according to their halo-centric distance (x-axis) and relative radial velocity (y-axis) for the more-massive neighbor exerting the largest tidal influence. Orbiting subhalos are represented by blue points and infalling by pink. Moving outwards, contours contain 15\%, 30\%, 60\%, and 90\% of the population. The orbiting halos fall within a triangle to the left of the figure. Infalling halos primarily fall outside this region with the exception of some objects that meet the orbiting velocity criteria but fell into their host less than half a dynamical time ago. The vertical black dotted line shows 1 $R_\text{vir}$, while the dashed curve shows the trajectory of a particle released at 2 $R_\text{vir}$ falling into a halo (see text for details). 
     }
     \label{fig:def3a}
\end{figure}

In this classification scheme, we split halos based on their history of interaction with a more-massive object. We adopt a similar strategy to \cite{Garcia_2023}, splitting our halos in accretion-time and velocity space. Accretion-time ($a_\text{acc}$) is defined here as the scale-factor at which the subhalo entered the virial volume of its host. By the criteria outlined below, we aim to separate subhalos that have completed their first pericentric passage, and thus have experienced substantial interaction with their host halo, from those which have yet to have such an interaction. We refer to these populations as orbiting and infalling respectively.

For halos not currently within the virial radius of a more-massive halo (HCD $ > 1$), a halo is orbiting if it has been more than half a dynamical time ($\tau_\text{dyn}$) since its accretion event ($a_\text{acc} < 0.87$). Otherwise, it is considered infalling.

For halos with HCD $\leq 1$, a halo is orbiting if:
1) it has been more $0.5\tau_\text{dyn}$ since its first accretion event ($a_\text{acc} < 0.87$) or 2) it has a positive radial velocity with respect to its host. Otherwise, it is considered infalling. Figure \ref{fig:def3b} shows the number density of halos in $a_\text{acc}$-velocity space and how this cut, represented by the white lines, separates the two populations. 

$0.5\tau_\text{dyn}$, or about one radius-crossing time, corresponds roughly to the timescale of a subhalo falling into its host and reaching pericenter. As such, it presents a natural and physically motivated cut between pre-pericenter and post-pericenter objects based on accretion time. By this accounting, there remains a significant population of halos (1.4 \%) that were accreted recently ($a_\text{acc} > 0.87$), but which have a positive radial velocity with respect to their host, suggesting they are post-pericenter. While, flybys may account for a small fraction of this population, the vast majority are likely objects that underwent pericentric passage in less than $0.5\tau_\text{dyn}$. The additional velocity cut was introduced to include this population among the halos classified as orbiting. 

Figure \ref{fig:def3a} shows where the populations of orbiting and infalling halos fall in relative velocity and position. There is substantial overlap between the two populations, particularly at low separations and negative velocities. The orbiting population is approximately symmetrical with respect to velocity. For comparison, we also show the trajectory of a massless particle falling from rest at 2 R$_\text{vir}$ into a \cite{Navarro_1997} mass profile with concentration ($\frac{R_\text{vir}}{R_s}$) of 10. From Figure \ref{fig:def3a}, the radial extent of the orbiting population is roughly traced by the massless particle trajectory above.

In their analysis of dark matter particles in simulations, \cite{Garcia_2023} also consider a third population: isolated particles or those not dynamically associated with a halo. By analogy, this would correspond to our isolated halos or those with no more-massive neighbors. In keeping with our previous decision to consider isolated halos as ‘centrals’ we likewise consider isolated halos here as ‘infalling,’ while acknowledging that such halos are not in fact dynamically associated with a larger halo.

\section{Methods}\label{sec:Methods}

\subsection{Optimal Isolation/Largest Neighbors Method}\label{subsec:Fixed}

As central/satellite classification is often performed based on an isolation criterion (e.g., no larger neighbor within $R_\mathrm{cut}$ projected kpc and $V_\mathrm{cut}$ km s$^{-1}$ redshift space), we first determine the optimum isolation criterion for separating centrals from satellites in our sample.  Generally, we expect that larger neighboring galaxies live in larger halos, which should have larger isolation radii.  Hence, a more general isolation criterion would be a function of the stellar mass of neighboring galaxies (e.g., no neighbor within $R_\mathrm{cut}(M_{*,\mathrm{neighbor}})$ projected kpc).  The performance of this type of method sets a lower limit to the accuracy we expect from the neural networks as described in the following sections.
 
 We begin by grouping galaxies by stellar mass in bins of width  $\Delta\log_{10} M_* = 0.25$ covering the range $\log_{10}(M_*/M_\odot) \in [9, 12]$. Then, we perform a search for each halo's nearest more-massive neighbors, selecting the closest neighbor in each stellar mass bin. A redshift separation threshold of $V_\mathrm{cut} = 2000$ km s$^{-1}$ is selected as it corresponds roughly with the virial velocity of a $\sim$$10^{14}~{\rm M}_\odot$ halo. This threshold choice helps retain physically associated systems—such as potential hosts or satellites—while reducing contamination from unassociated galaxies projected along the line of sight. The virial radius $R_\text{vir}(M_*)$ is taken to be the average virial radius of objects in the stellar mass bin $M_*$.

We express the cut-off radial separation ($R_\text{cut}$) between central and satellite as an analytical expression of the form $R_\text{cut} = a\cdot R_\text{vir}(M_*)^b + c$.  A halo is classified as a satellite if it has a neighbor at $R_\text{sep}< R_\text{cut}(M_*)$ where $M_*$ is the stellar mass of the neighbor. The values of $a$, $b$, and $c$ are tuned to minimize the fraction of miscategorized halos. The optimized expression is: 

\begin{equation}
    \frac{R_{\text{cut}}}{1 \text{Mpc}} = 0.75\left(\frac{R_\text{vir}}{1 \text{Mpc}} \right)^{0.56} - 0.14.
    \label{equation:r_eff}
\end{equation}

\begin{figure}
     \centering
     \includegraphics[width=0.45\textwidth]{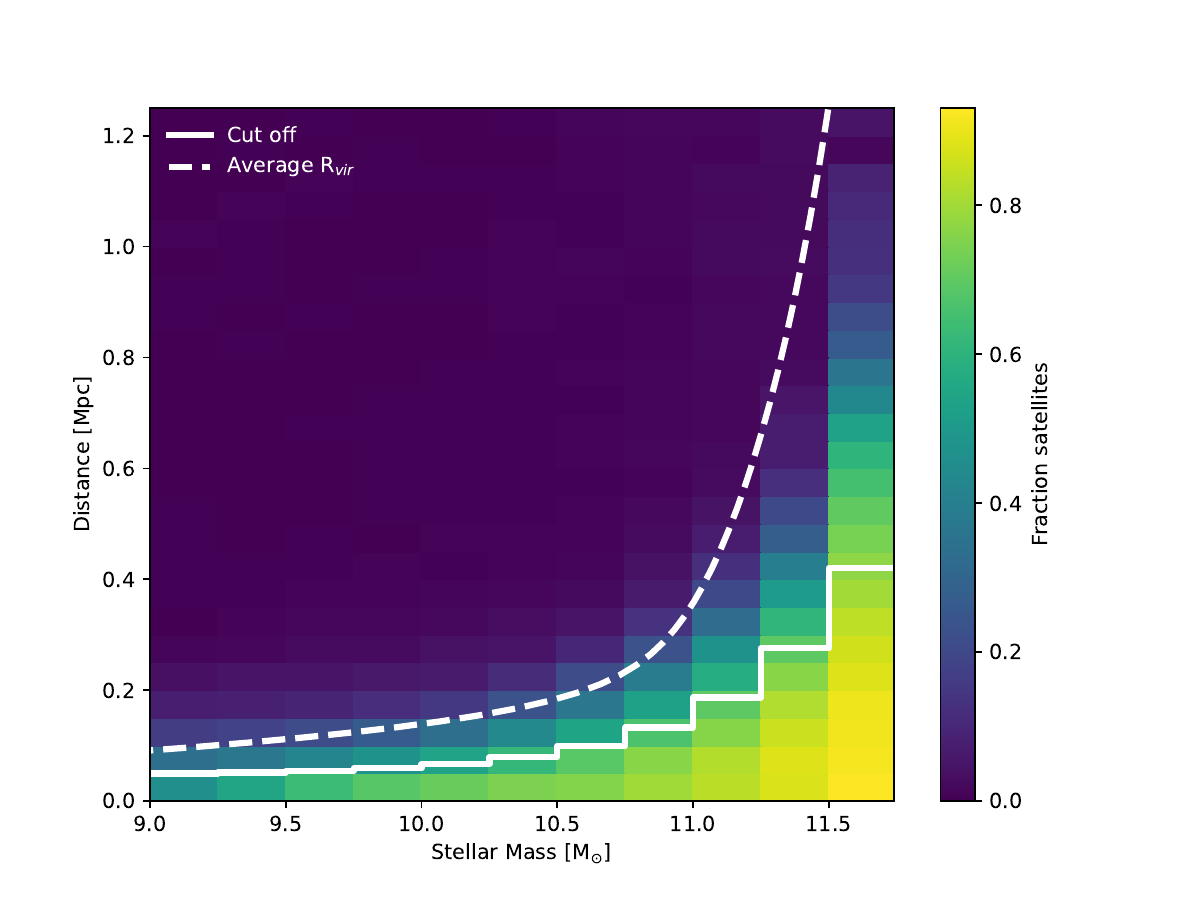}
     \caption{For each object, we consider its closest larger neighbor scaled by the cut-off criteria (see text for full description of how this is determined). Bins in stellar mass of neighbor and projected separation are colored according to the fraction of objects in that bin that are satellites. The solid white histogram shows the cut made to determine satellites status as defined by Equation \ref{equation:r_eff}. Objects with a more massive neighbor falling below that line are classified as satellites. For reference, the dashed white line shows the average virial radius (not projected) for a neighbor at that stellar mass. The cut-off we derive falls near the bins for which 50\% are satellites (optimizing accuracy).  In contrast, using a cut-off at the average R$_{\text{vir}}$ for a given neighbor galaxy's stellar mass would increase completeness at substantial cost to purity.}
     \label{fig:manual}
\end{figure}

To illustrate the fraction of satellites enclosed by this criteria, we consider all potential neighbors for each object and compute

\begin{equation}
    D = R_\text{sep} - R_\text{cut}(M_*^\text{neighbor}).
\end{equation}
We then select, for each object, the neighbor with the smallest value of $D$. This is the object's `nearest' neighbor when scaled by the cut-off criteria. We then create a 2D histogram in the space of neighbor stellar mass versus separation from the selected neighbor in Figure \ref{fig:manual}. Each bin in this histogram is colored by the fraction of objects in that bin that are satellites. This representation reveals how satellite likelihood varies with both the mass and proximity of the closest neighbor.

The white histogram shows the cut-off criteria, here calculated at the median stellar mass value for each bin. All objects with neighbors falling below the cut-off line are classified as satellites. We note that the optimal cut-off line lies substantially below the average virial radius for a given stellar mass bin, especially at higher stellar masses, due to contamination from projection effects. When applying the cut-off criteria to the test set of galaxies, no such binning by stellar mass is applied---i.e., $R_\mathrm{vir}(M_\ast)$ is interpolated between bins.

The same approach is applied for classifying historical centrals versus satellites and infalling versus orbiting populations. For all three classification cases we independently optimize the expression for $R_\text{cut}$ so as to minimize the miscategorization fraction. We find a similar cut, but one that scales more strongly with $R_\mathrm{vir}$, is optimal in the case of historical centrals vs.\ satellites and infalling versus orbiting populations compared to that shown in Eq.\ \ref{equation:r_eff} for present day centrals versus satellites. The details of these expressions are left to their respective subsection in Section \ref{sec:Results}.

\subsection{Neural Network}\label{subsec:Network}

With our neural network approach, we constructed input vectors that include a broad set of potentially relevant environmental features. Rather than manually selecting a minimal set of physically motivated inputs, we allow the network to learn which aspects of the environment are most informative by adjusting the weights during training. In this way, the network is capable of extracting useful patterns from a richer, more flexible description of the local environment than used in the optimal isolation method described in the previous section.

Each input vector to the neural network is designed to describe both the target galaxy and its local environment. We characterize the local environment using the stellar masses, projected positions, and velocities of the galaxy’s nearest neighbors. Specifically, we define a galaxy’s neighbors as those with the smallest projected separations that satisfy two criteria: (i) a redshift separation of less than 2000 km s$^{-1}$ and (ii) a stellar mass no more than 1.5 dex below that of the target galaxy. We adopt a projected-space definition to enable seamless application to observational catalogs, where full 3D positions are typically unavailable.

The redshift separation threshold of 2000 km s$^{-1}$ is the same applied in the isolation criteria, corresponding approximately with the virial velocity of a $\sim$$10^{14}~{\rm M}_\odot$ halo. The stellar mass cut prevents nearest-neighbor selections from being dominated by low-mass satellites in dense environments, allowing a more representative sampling of the local halo context.

We explored two methods for encoding this environmental information in the input vectors: (i) the properties of the $k$ nearest neighbors and (ii) galaxy counts within cylindrical apertures. We find that the $k$-nearest neighbors approach yields substantially better predictive performance, and thus adopt it for the remainder of this study. Further details on the counts-in-cylinders method are provided in Appendix~\ref{sec:AppA}.

\subsubsection{$k$ Nearest Neighbors}\label{subsec:kNN}

For the $k$ Nearest Neighbors ($k$NN) environment description, we create an array of the target galaxy's $k$ nearest neighbors, where $1 < k < 50$. Each neighbor has three values associated with it --- projected separation from the target, redshift separation, and a relative stellar mass ranking. The neighbors are sorted in order of projected distance to create a $k\times 3$ vector. In addition, we include the relative stellar mass ranking of the target galaxy itself in the input vector.

\subsubsection{Network Structure \& Training}

For each of the three classification tasks described in Section \ref{sec:Def}, we construct a feed-forward neural network using the \texttt{PyTorch} framework \citep{pytorch}. The network takes $3k + 1$ $k$NN input features, where $k$ is the number of neighbors included. In our final implementation, we adopt $k = 25$, resulting in an input vector of length 76.

All input features are normalized to have zero mean and unit variance prior to being passed through the network. The architecture of the network consists of an initial hidden layer with 250 units, followed by additional fully connected layers whose sizes decrease by a factor of 0.9 relative to the preceding layer. A dropout rate of 0.06 is applied after each hidden layer to mitigate overfitting. The final layer outputs a scalar $\mathcal{L} \in [0, 1]$, representing the predicted likelihood that the galaxy is a satellite.

The network is trained using the binary cross-entropy loss function,

\begin{equation}
\mathcal{L} = -\frac{1}{N}\sum_{i=1}^N(y_i \log{p_i} + (1-y_i)\log{(1-p_i)})
\label{eq:BCE}
\end{equation}

where $y_i \in {0,1}$ is the true class label for instance $i$, and $p_i$ is the network’s predicted probability that $y_i = 1$. Galaxies with predicted likelihoods $\mathcal{L} \leq 0.5$ are classified as centrals, while those with $\mathcal{L} > 0.5$ are classified as satellites.

To determine appropriate values for key hyper-parameters—including the number of neighbors $k$, the number and sizes of hidden layers, and the learning rate—we performed an automatic optimization using the \texttt{Optuna} framework \citep{optuna}, which finds the best option out of a user-specified hyperparameter range, selecting for minimum validation after 50 epochs. Table~\ref{tab:Fixed} summarizes the range of hyper-parameters considered and the final values adopted. Although we also experimented with including convolutional layers prior to the fully connected layers for feature extraction, we found no significant improvement in performance and thus did not include them in the final model. As information on the full 25 nearest neighbors may not be possible in various observational scenarios (e.g., survey boundaries), we explored the impact of using fewer than 25 neighbors, finding that dropping to as few as 5 neighbors has only a small impact on the classification accuracy for current centrals versus satellites. Appendix \ref{sec:AppD} further outlines the performance of the network with fewer than 25 neighbors.

Each of the three networks, corresponding to the three classification tasks, was optimized independently. However, the resulting optimal hyper-parameters were highly consistent across all cases. For simplicity and reproducibility, we adopt a unified set of hyper-parameters for all three networks. After fixing these parameters, each model is retrained on the SMDPL dataset for an additional 100 epochs to ensure convergence, with the final network weights selected from the epoch with the lowest validation loss. Model performance is then evaluated using the Bolshoi-Planck dataset, which was held out from training and tuning.

\begin{table}[h]
 \caption{Neural network parameters}
 \centering
 \begin{tabular}{llc}
  \hline
  Parameter & Values Considered & Final Values\\
  \hline
  Number of Neighbors & $x\in (1,50)$ & 25\\
  Learning Rate &  $x\in (10^{-5}, 0.001)$ & 0.0005\\
  Hidden Layers & $x\in (2,6)$ & 3\\
  Hidden Dimension & $x\in (100,300)$ & 300\\
  Narrowing Factor & $x\in (0.5,1.0)$ & 0.9\\
  Dropout Rate ($f$) & $x\in (0.01, 0.1)$ & 0.06\\
  Batch Size & 32, 64, 128, 256 & 256\\
  Optimization Function & Adam, SGD & Adam\\
  Activation Function & ReLU, sigmoid, tanh & ReLU\\
  \hline \\
   \multicolumn{3}{c}{\parbox{0.9\columnwidth}{\footnotesize \textit{Notes:} Abbreviations are as follows: stochastic gradient descent (SGD) and rectified linear unit (ReLU). Narrowing factor is defined as the fractional number of nodes in a given hidden layer of the network when compared to the previous hidden layer.}}
 \end{tabular}
 \label{tab:Fixed}
\end{table}

\subsection{Performance Metrics}\label{subsec:Metrics}

For each classification case presented in Section \ref{sec:Def}, we assume all objects belong to one of two classes. In each classification case and for each method, we first consider the overall classification accuracy. This is defined as the fraction of all objects that are assigned the correct class. We further consider the purity or the recovered sample for a given class, which provides the ratio of objects correctly assigned to a given class to the total number of objects assigned to the class.
We also calculate the completeness of the sample, that is, the fraction of objects of a given class that are recovered.

\section{Classification Performance}\label{sec:Results}

\subsection{Current Centrals vs. Satellites}\label{subsec:CurrentCvS}

Central galaxies, as defined in Section \ref{subsec:Def1}, constitute $\sim 75\%$ of all objects with stellar mass $M_* > 10^9 \Msun$ in the test galaxy-halo catalogue. This fraction increases with stellar mass and sets a lower bound on overall classification accuracy --- a naive model that labels all galaxies as centrals would be accurate 75\% of the time. 

Applying the optimal isolation criteria outlined in Section \ref{subsec:Fixed} and defined in Equation \ref{equation:r_eff}, we recover the true central/satellite classification for 86.5\% of objects in the test set. Among the misclassified galaxies, $\sim$35\% are centrals incorrectly identified as satellites, while $\sim$65\% are satellites misidentified as centrals. A flat increase in the value of $R_\text{cut}$ shifts this misclassification bias towards the former but decreases the overall accuracy.

The trained neural network, as described in Section \ref{subsec:Network}, reduces the misclassification fraction by 24\%, correctly classifying 89.7\% of objects. Of the $\sim$10\% misclassified, $\sim$60\% are centrals mislabeled as satellites and $\sim$40\% are satellites mislabeled as centrals. In cases where one values purity in the centrals sample over completeness or wishes to remove the bias in the class populations, the network can be re-trained with a higher weight put to satellites than centrals (2:1). Doing so reduces the completeness of the recovered central population from 92\% to 88\%, but increases the purity of the sample from 94\% to 97\%. These results are consolidated in Table \ref{tab:Results}. Appendix \ref{sec:AppB} presents an analysis of how varying class weights during training affects classification performance.

\begin{figure}
\centering
\includegraphics[width=0.45\textwidth]{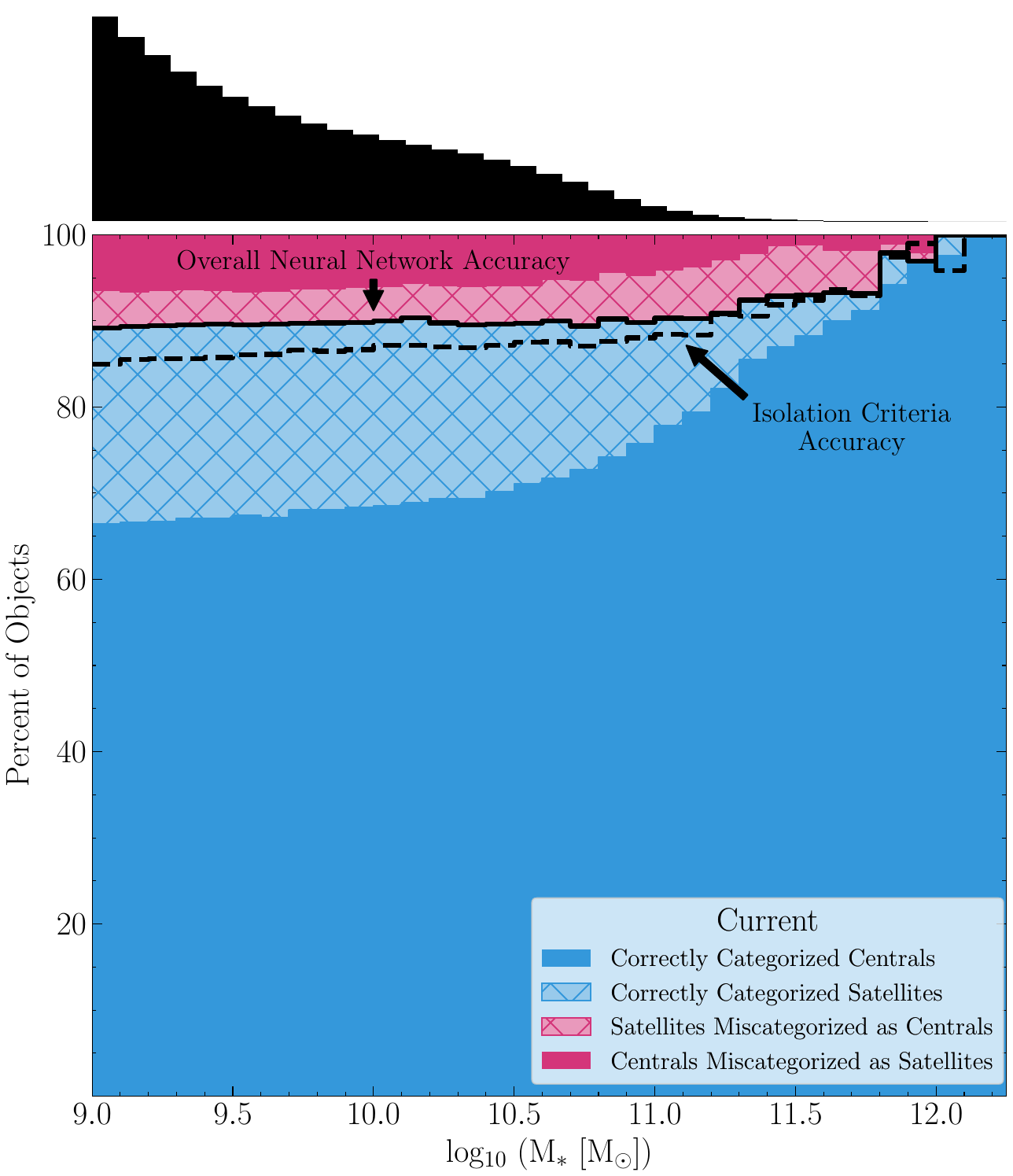}
\includegraphics[width=0.45\textwidth]{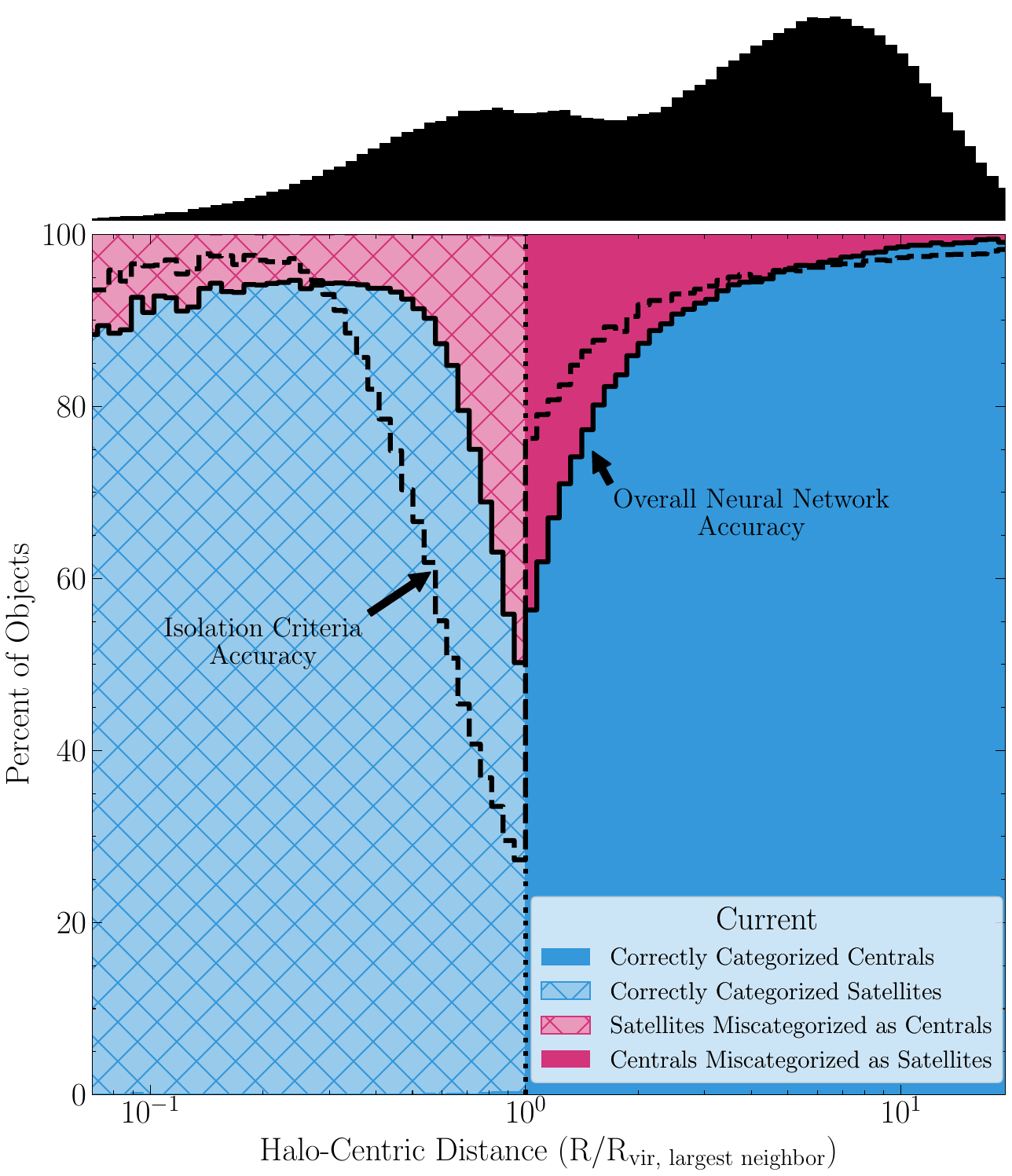}
\caption{The performance of the network classifying galaxies as current centrals of satellites and the number density of halos in the sample (black histograms) is shown as a function of stellar mass (Top) and halo-centric distance (Bottom). Galaxies are broken into four categories based on both their true and predicted classification as represented by the colored regions, with correctly classified galaxies shown in blue and incorrectly classified in pink. The overall percentage of galaxies that are correctly classified by the network is represented by the black solid line. The dashed line shows the overall percentage correctly classified by the isolation criteria for comparison. The dotted vertical line in the bottom plot illustrates the dividing line between current central and satellite as described in Section \ref{subsec:CurrentCvS}.
}
\label{fig:def1_acc}
\end{figure}

Figure \ref{fig:def1_acc} shows the accuracy of the isolation criterion (dashed line) and neural network (solid line) as a function of the stellar mass of the target galaxy (Top) and minimum halo-centric distance (Bottom). The colored regions in Figure \ref{fig:def1_acc} show the cumulative percentage of objects in the test set that are correctly classified by the neural network (blue) or misclassified (pink), further broken into true centrals (solid) and true satellites (hatched). Reducing the area of the pink hatched region, for example, would be to reduce the number of satellites mislabeled as centrals, and thus improve the purity of the central sample. These regions are included as guides to which misclassification type dominates the network results at different scales. The histograms above each plot indicate the distribution of objects in our test-set.

Figure \ref{fig:def1_acc} (Top) illustrates that for both the isolation criteria and the neural network, classification accuracy improves with increasing stellar mass, reflecting the rising fraction of central galaxies in that regime. The high central fraction at large stellar masses means that these objects can generally be classified confidently without the use of any environmental information, though the regime of $M_* > 10^{10.5} M_{\odot}$ represents a small fraction of the overall galaxy population in our sample (see histogram). In contrast, at lower masses where the majority of our galaxy population can be found, environmental information is essential for separating low-mass isolated objects from satellites.

Figure \ref{fig:def1_acc} (Bottom) shows the fraction of objects by category as a function of the minimum halo-centric distance ($R/R_\text{vir, largest neighbor}$). If the halo-centric distance between an object and its neighbor is less than one (to the left of the vertical black line), this indicates that the object's current position falls within the virial radius of the neighbor, and thus is a satellite by our definition. The sharp dip in classification accuracy around this boundary for both methods is expected as uncertainty in the halo masses of these objects ($\gtrsim$0.2 dex when estimated from stellar mass rankings), as well as projection effects, will result in objects being scattered to either side of the boundary.

For the isolation criteria method, accuracy begins declining sharply at lower values of $R/R_\mathrm{vir,largest}$ than the neural network method, with a sharp jump at $R/R_\mathrm{vir,largest} = 1$, continued by a higher accuracy regime slightly beyond the virial radius. This suggests that the the isolation cut as implemented requires a satellite target's host to be relatively close for the target to be correctly labeled, while target's at large separations from more-massive neighbors are classified as centrals. Shifting the isolation threshold to larger radii would reduce the contamination of the central pool with mislabeled satellites, which is driving this sharp decline at low separations, but at the cost of increasing the number of centrals misclassified as satellites at larger separations.

\begin{figure}
\centering
\includegraphics[width=0.45\textwidth]{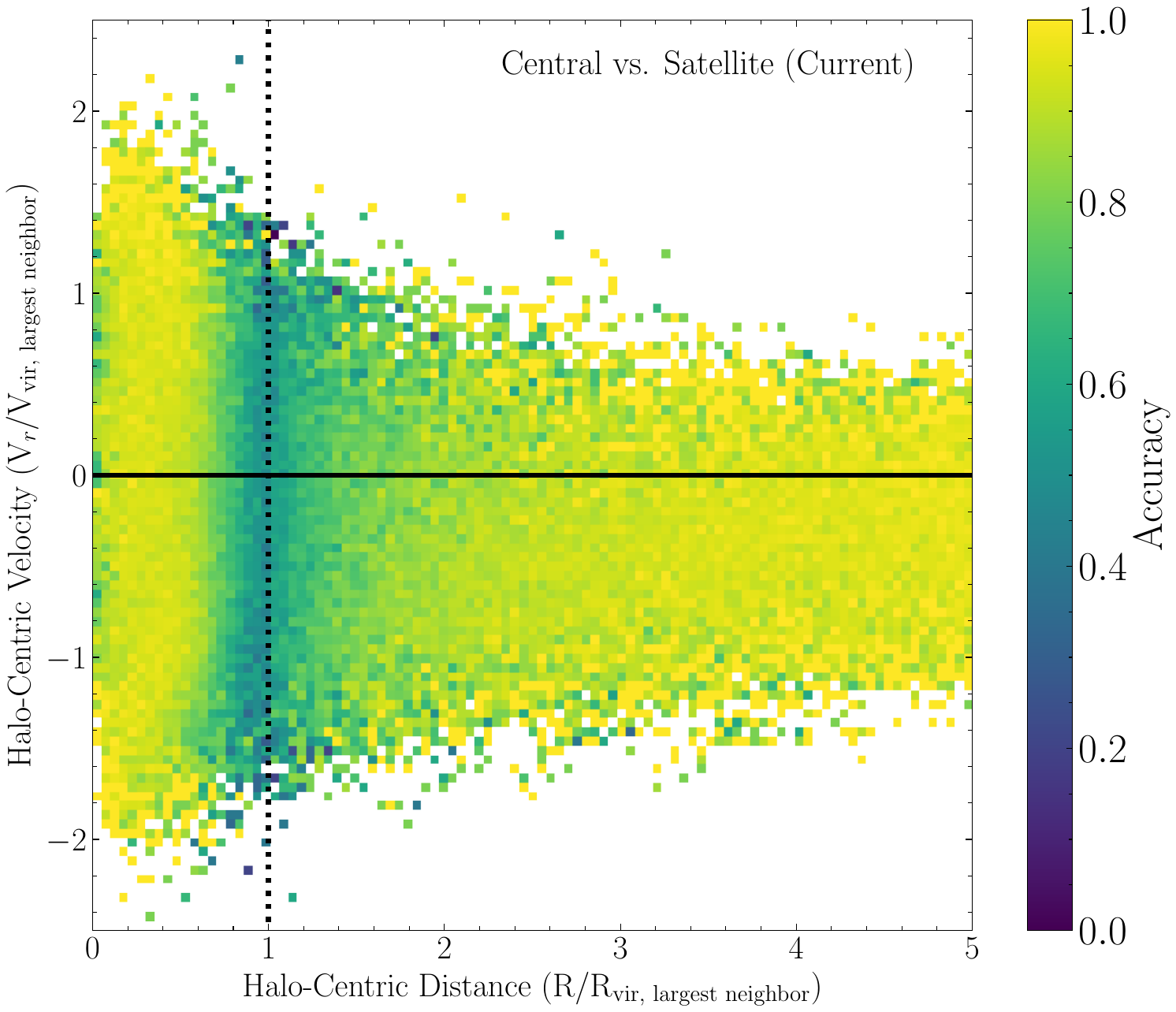}
\caption{The average classification accuracy of the network is shown in bins of the relative position and velocity of a halo to its neighbor with the largest tidal influence. As in Figure \ref{fig:def1}, the vertical dotted line represents the true boundary between satellite and central. The lowest accuracy can be found in the region of R/R$_{\text{vir, largest neighbor}} = 1.0\pm0.25$, with no clear velocity dependence.
}
\label{fig:def1_ps}
\end{figure}

The current central/satellite division is dependent only on the mass and positions of the halos at the current snapshot regardless of the halos’ histories or relative velocities. Therefore, as illustrated by Figure \ref{fig:def1_ps}, the relative velocity between the halo and its most tidally-influential neighbor does not substantially impact the accuracy of the network under this classification scheme.

The two modes which dominate misclassification can be generally categorized as 1) uncertainties in redshift space separations and 2) uncertainties in halo masses. To determine which failure modes are important in different regimes, we tested removing the uncertainty in redshift-space positioning by providing the network with 3D separations between objects (not available in observations) rather than 2D + velocity separation. The details of this method can be found in Appendix \ref{sec:AppC}. We find that approximately 60\% of misclassifications are due to projection effects. The remaining misclassifications are composed primarily of objects falling near the hard boundary between central and satellite at the virial radius. In these cases, even with the removal of projection effects, uncertainty in halo mass can scatter an object to either side of the boundary, or, in a few cases where two objects have similar stellar masses, result in the swapping of central and satellite labels.

We additionally analyzed how the network accuracy corresponded to the reported prediction confidence of the network. The network's output is a likelihood value between 0 and 1, with 0 representing overwhelming information in the input suggesting the object is a central and 1 representing a confident prediction that the object is a satellite. The confidence of the classification of a given object is represented by the distance of the predicted value from 0.5 (i.e., no information preferring one classification over the other). 

We find that trends in increased confidence correspond well with trends in accuracy over both stellar mass and halo-centric distance. This indicates that the network is correctly identifying these objects as being near the boundary between classes or having inputs with insufficiently identifying information and subsequently assigning them classes with low confidence. In a similar manner as adjusting class weights, a more pure sample of centrals could be found by increasing the confidence level required to assign an object as a central (e.g., changing the likelihood cut-off from $\mathcal{L}\leq 0.5$ to $\mathcal{L} \leq 0.3$), though this also has a substantial impact on the size of the central sample recovered (see Appendix \ref{sec:AppB}).

 \subsection{Satellite History}\label{subsec:HistResults}

\begin{figure}
\centering
\includegraphics[width=0.45\textwidth]{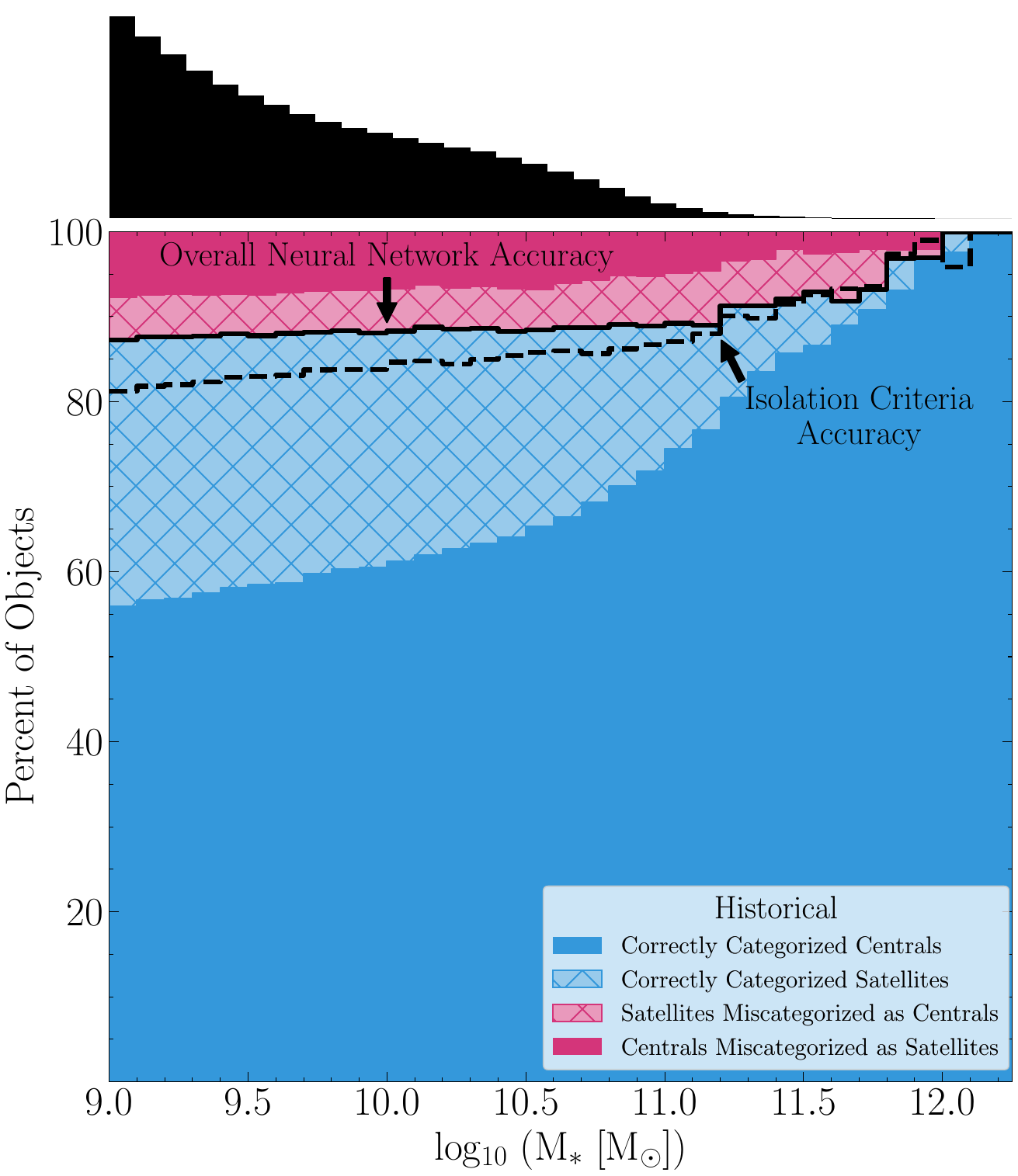}
\includegraphics[width=0.45\textwidth]{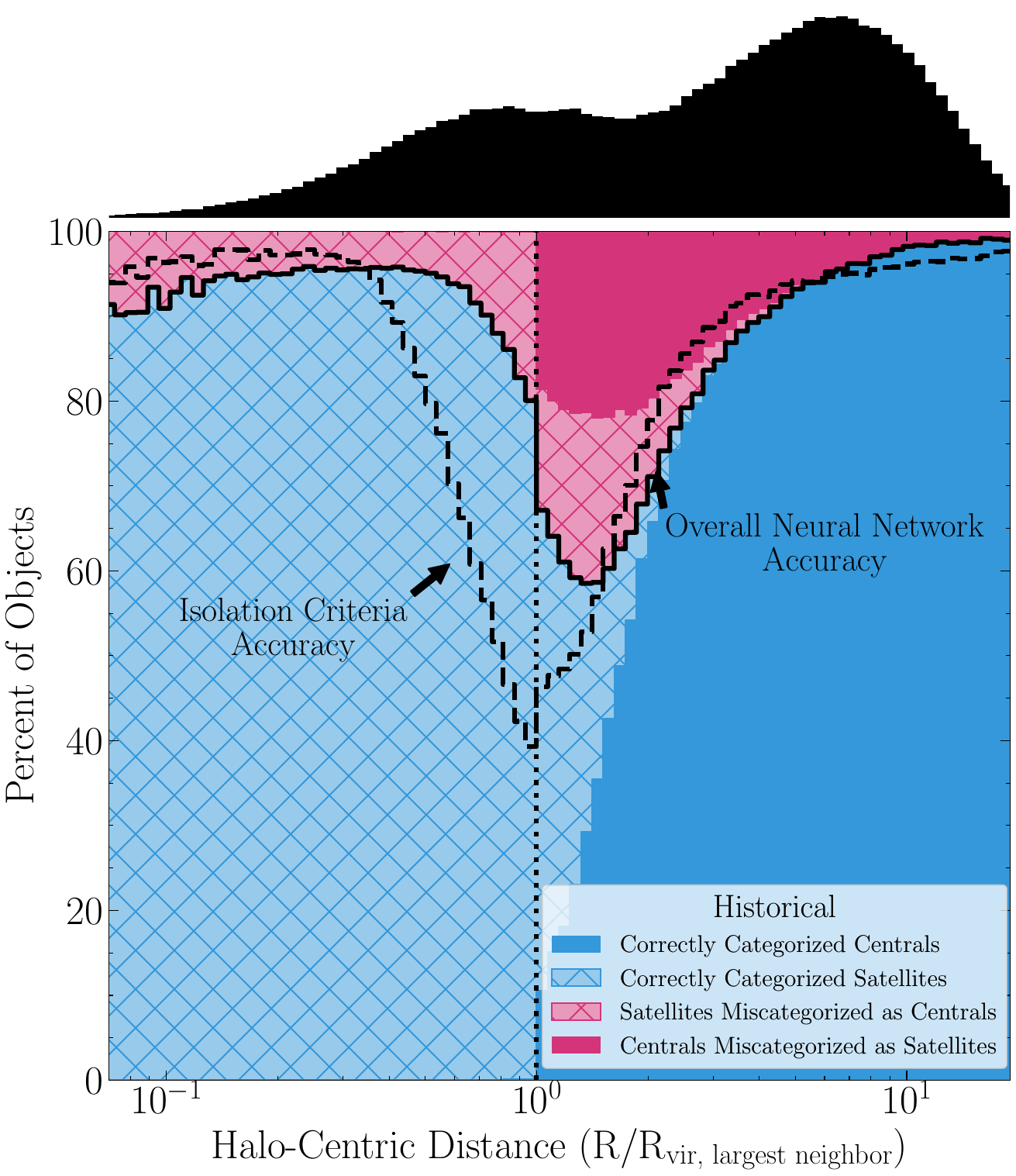}
\caption{The equivalent of Figure \ref{fig:def1_acc} is shown for the performance of the network on classifying halos into historically centrals or satellites and the number density of halos in the sample (black histograms) as a function of stellar mass (Top) and halo-centric distance (Bottom). Galaxies are broken into four categories based on both their true and predicted classification as represented by the colored regions, with correctly classified galaxies shown in blue and incorrectly classified in pink. The overall percentage of galaxies that are correctly classified by the network is represented by the black solid line. The dotted vertical line in the bottom plot shows 1 $R_\mathrm{vir}$, or what we consider the halo boundary, for most tidally-influential neighbor.
}
\label{fig:def2_acc}
\end{figure}

As described in section \ref{subsec:Def2}, objects are defined as having historically been a satellite if they satisfy the satellite criteria in the current snapshot or during any previous snapshot. By this definition, the optimal isolation criteria ($R_\text{cut}$; sec. \ref{subsec:Fixed}) is

\begin{equation}
\frac{R_{\text{cut}}}{1 \text{Mpc}} = 0.86\left(\frac{R_\text{vir}}{1 \text{Mpc}} \right)^{0.61} - 0.12.
\label{equation:r_eff2}
\end{equation}

Using this criteria, we recover the historical central versus satellite classification of our test dataset at an overall accuracy of 83.7\%. Using a neural network trained to make this classification, we are able to reduce the misclassification rate over the optimal isolation criteria and recover the true classification of $\sim$88\% of objects in our test dataset. Approximately 58\% of the network’s misclassifications are centrals miscategorized as historically satellites, with the remaining 42\% being historical satellites classified as never having been a satellite.

The misclassification rate of the neural network follows the same general trend with stellar mass as for the current central vs.\ satellite classification (Figure \ref{fig:def2_acc}; Top). However, the rate of misclassification for historical satellite status peaks at larger halo centric distances (Figure \ref{fig:def2_acc}; Bottom). Under this classification scheme, the physical separation between a halo and its most tidally-massive neighbor represents a gradual transition between the two classes, rather than a sharp cut, as both classes contain objects with $R/R_{\text{vir, largest neighbor}} > 1.0$. This overlap between the two populations shifts the area of greatest uncertainty to higher values of $R/R_{\text{vir, largest neighbor}}$, where, if given access to the full phase-space information, the relative velocity of the object and the satellites would inform us as to whether the object was moving towards its potential host (and thus possibly on first infall) or moving away from its potential host (and thus was likely a satellite of said host in a previous snapshot).

\begin{figure}
\centering
\includegraphics[width=0.45\textwidth]{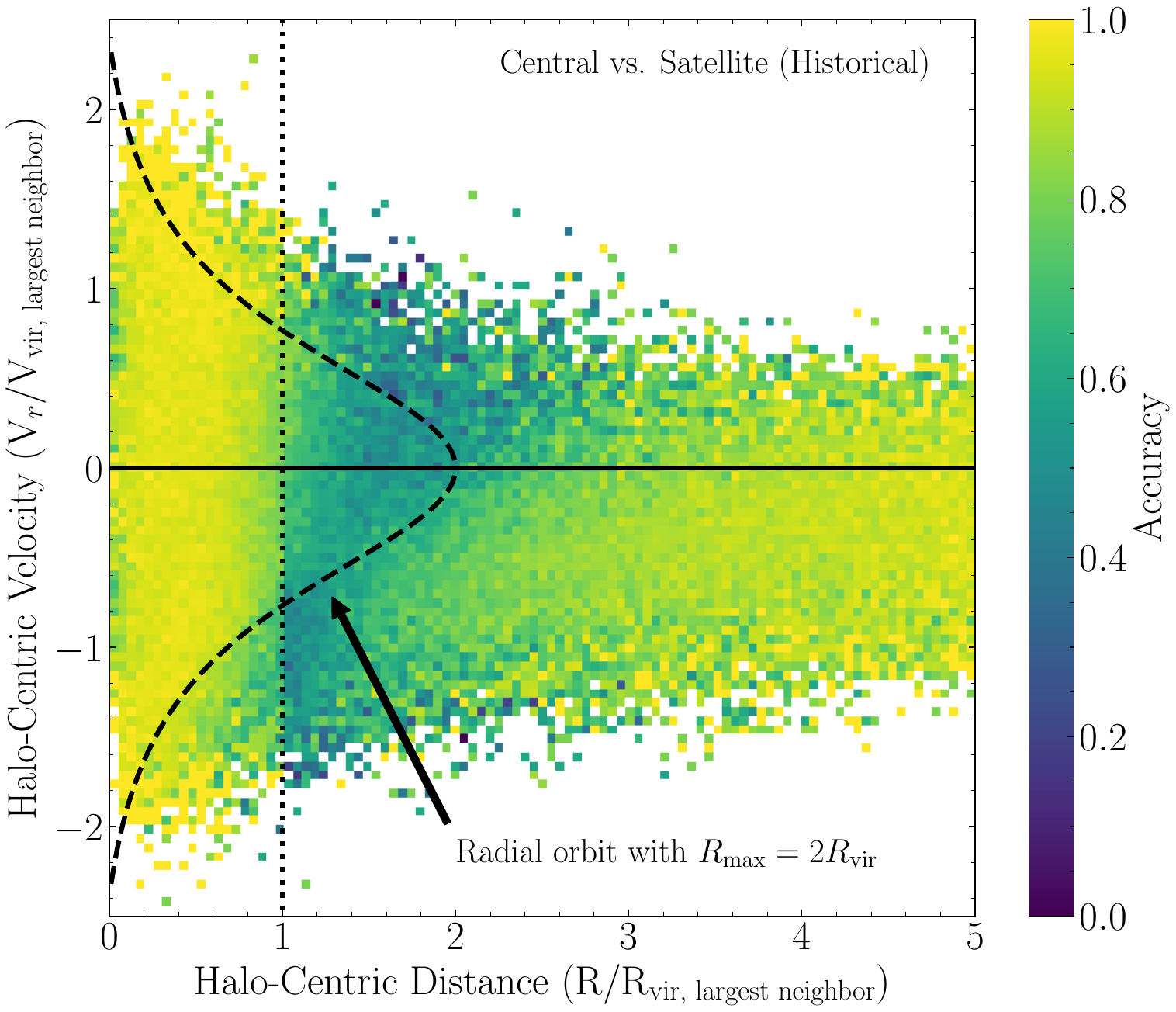}
\caption{The overall accuracy of the network in recovering both historical centrals and satellites is shown as a function of the relative position and velocity of a halo to its neighbor with the largest tidal influence. The vertical dotted line represents 1 $R_{\text{vir}}$ for the most tidally-influential neighbor. As in Figure \ref{fig:def3a}, the black dotted curve shows the trajectory of a massless particle released at 2 $R_\mathrm{vir}$ falling into a halo. The lowest accuracy region has shifted from the region surrounding the 1 $R_{\text{vir}}$ boundary to a region at $> 1 R_\mathrm{vir}$ from its largest neighbor with evidence of a velocity dependence roughly following the trajectory shown in the dotted black curve.
}
\label{fig:def2_ps}
\end{figure}

While the neighbor information provided to the network provides some clues as to the history of the halo, as seen in the relatively high classification accuracy, there remains ambiguity as to the halo’s relative velocity to its neighbors. This ambiguity is reflected in the trend of accuracy of the network as a function of relative velocity as shown in Figure \ref{fig:def2_ps}. For current satellites (those to the left of the vertical dashed line), accuracy is largely independent of velocity. However, for former satellites or objects on first infall, the network struggles to distinguish between those entering and those exiting the virial volume of a neighbor.

\begin{table*}
 \caption{Classification Performance}
 \centering
 \begin{tabular}{c c c c c}
  \hline
  \textbf{Case} & \textbf{Method} & \textbf{Overall Accuracy} & \textbf{Purity (Central/Infalling)} & \textbf{Completeness (Central/Infalling)} \\
  \hline
  \multirow{3}{*}{\parbox{3cm}{\centering Current Central vs. Satellite}} 
   & Optimal Isolation & 86.5\% & 88.7\% & \textbf{93.9\%}\\
   & Neural Network (Default) & \textbf{89.7\%} & 94.2\% & 91.9\%\\
   & Neural Network (Weighted 2:1) & 89.0\% & \textbf{96.8\%} & 88.1\%\\
  \hline
  \multirow{2}{*}{\parbox{3cm}{\centering Historical Central vs. Satellite}} 
   & Optimal Isolation & 83.7\% & 84.4\% & \textbf{93.2\%}\\
   & Neural Network & \textbf{88.2\%} & \textbf{92.4\%} & 89.9\%\\
  \hline
  \multirow{2}{*}{\parbox{3cm}{\centering Infalling vs. Orbiting}} 
   & Optimal Isolation & 82.6\% & 84.6\% & \textbf{92.5\%}\\
   & Neural Network & \textbf{86.5\%} & \textbf{91.4\%} & 89.4\%\\
  \hline
  \multicolumn{5}{p{\textwidth}}{\footnotesize \textit{Notes:} Central/infalling purity and completeness refer to the purity and completeness of the recovered sample of the class corresponding to current centrals, historical centrals, and infalling objects in the three classification cases considered. These performance metrics are defined in Section \ref{subsec:Metrics}. The highest performance by each metric is shown in bold.}
 \end{tabular}
 \label{tab:Results}
\end{table*}

 \subsection{Orbiting vs. Infalling}

 \begin{figure}
     \centering
     \includegraphics[width=0.45\textwidth]{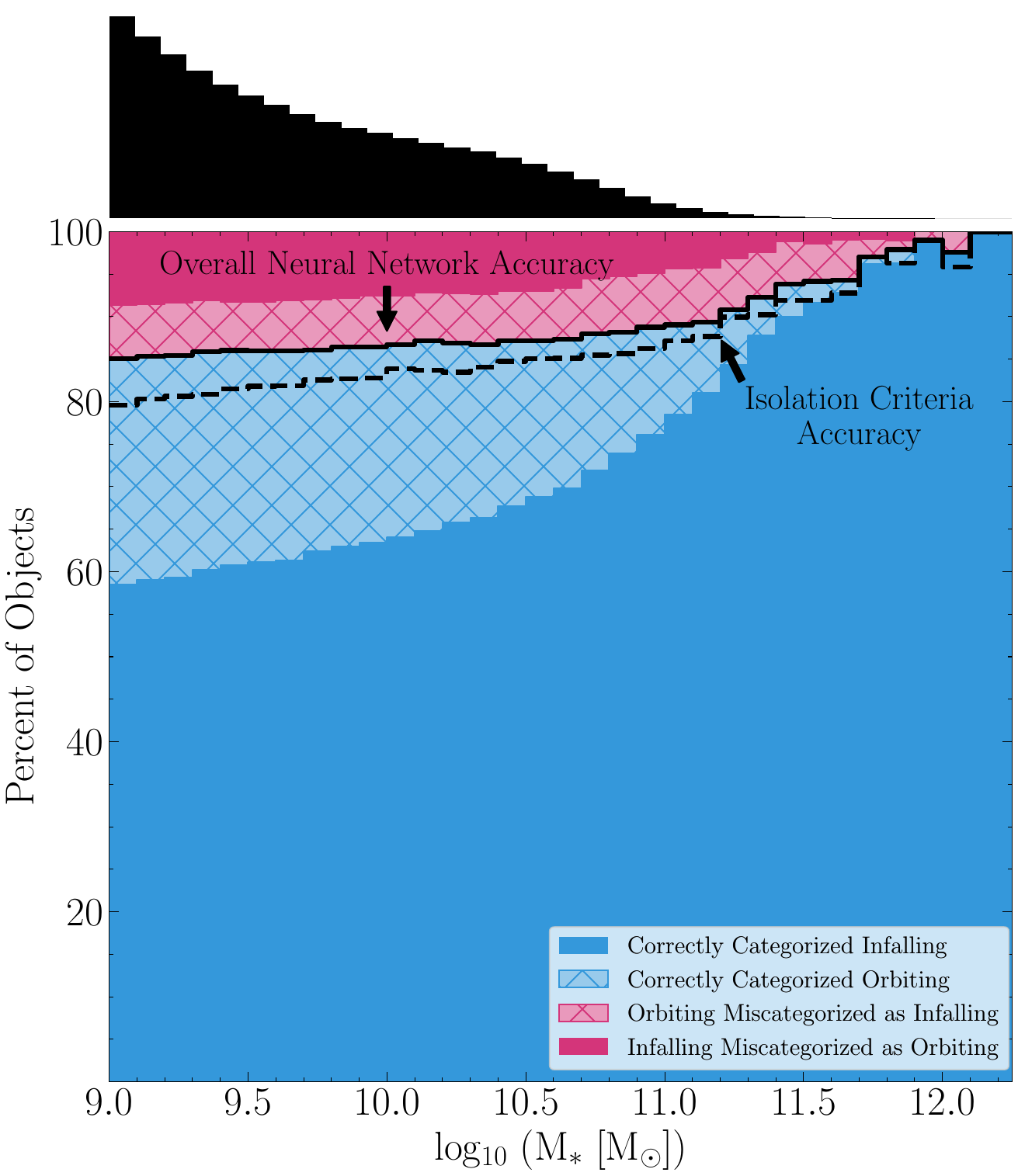}
     \includegraphics[width=0.45\textwidth]{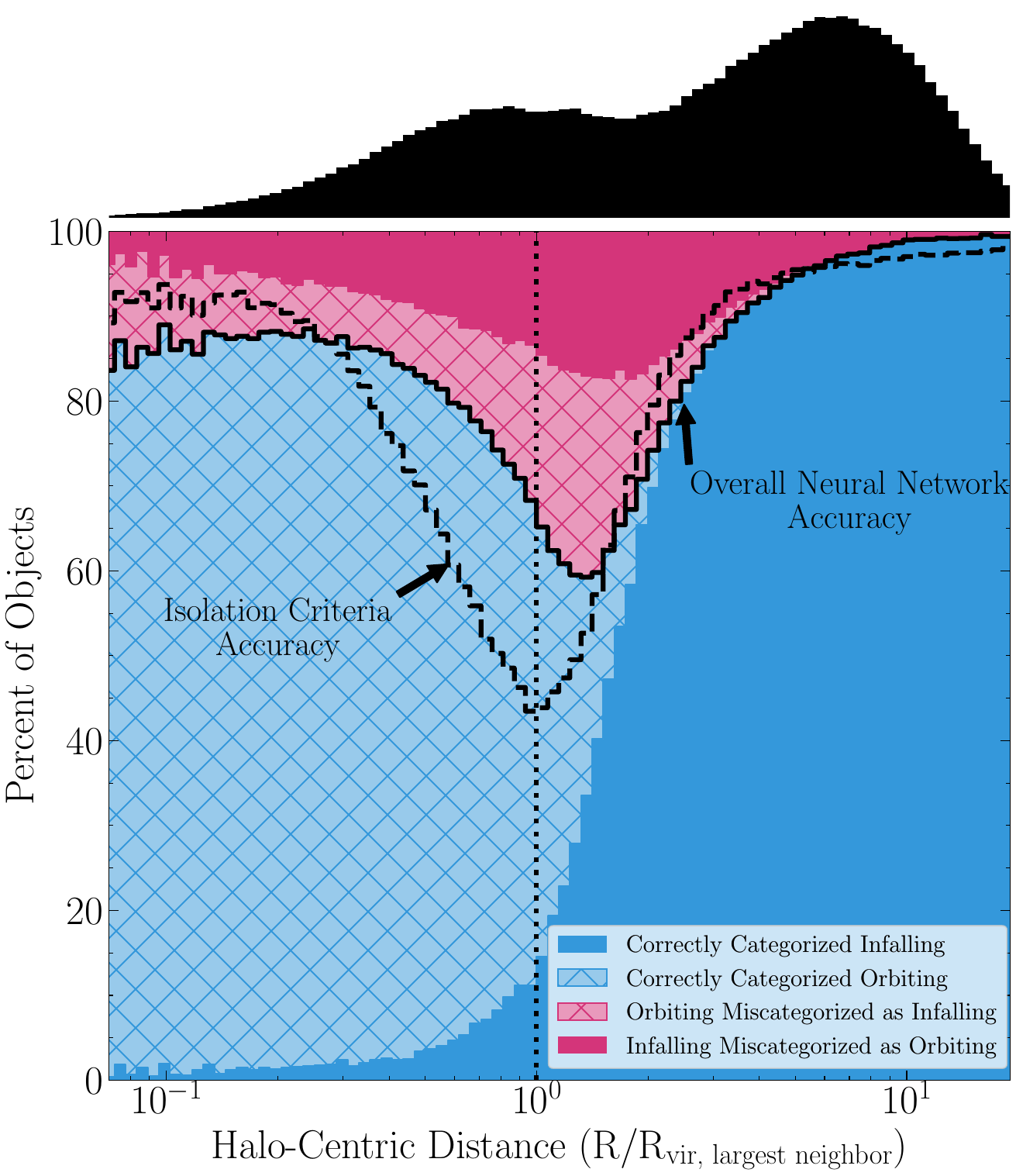}
     \caption{The equivalent of Figures \ref{fig:def1_acc} and \ref{fig:def2_acc} is shown for the performance of the network on classifying halos into infalling and orbiting populations and the number density of halos in the sample (black histograms) as a function of stellar mass (Top) and halo-centric distance (Bottom). Galaxies are broken into four categories based on both their true and predicted classification as represented by the colored regions, with correctly classified galaxies shown in blue and incorrectly classified in pink. The overall percentage of galaxies that are correctly classified by the network is represented by the black solid line. The dotted vertical line in the bottom plot shows 1 $R_{\text{vir}}$, or what we consider the halo boundary, for most tidally-influential neighbor.
     }
     \label{fig:def3_acc}
\end{figure}

In this section, we aimed to separate the halo population into objects that are currently in orbit to a more massive object and those that are infalling. The orbiting versus infalling distinction is complicated by lack of access to the full 3D phase space for each object in the data vector we provided to the network.

By the isolation criteria method we find an expression for $R_\text{cut}$ given by Eq. \ref{equation:r_eff3}:

 \begin{equation}
    \frac{R_{\text{cut}}}{1 \text{Mpc}} = 0.85\left(\frac{R_\text{vir}}{1 \text{Mpc}} \right)^{0.63} - 0.14.
    \label{equation:r_eff3}
\end{equation}

Using the optimal isolation cut, we recover the true orbiting/infalling class in 82.6\% of cases. As in the previous sections, the trained neural network improves upon the performance of the optimal isolation cut. In this case, it recovers the true classification 86.5\% percent of the time (see Table \ref{tab:Results} for details). This is a lower recovery rate than the previous two definitions, reflective of the importance of the missing velocity information to this particular classification. As seen in Figure \ref{fig:def3_acc} (bottom), this misclassification is far more common for objects at smaller physical separation from the neighbor exerting the largest tidal influence, including those that have recently become satellites. A trend towards lower misclassification rate with higher stellar mass can be seen in Figure \ref{fig:def3_acc} (top) as was found with the previous classification cases.

 \begin{figure}
     \centering
     \includegraphics[width=0.45\textwidth]{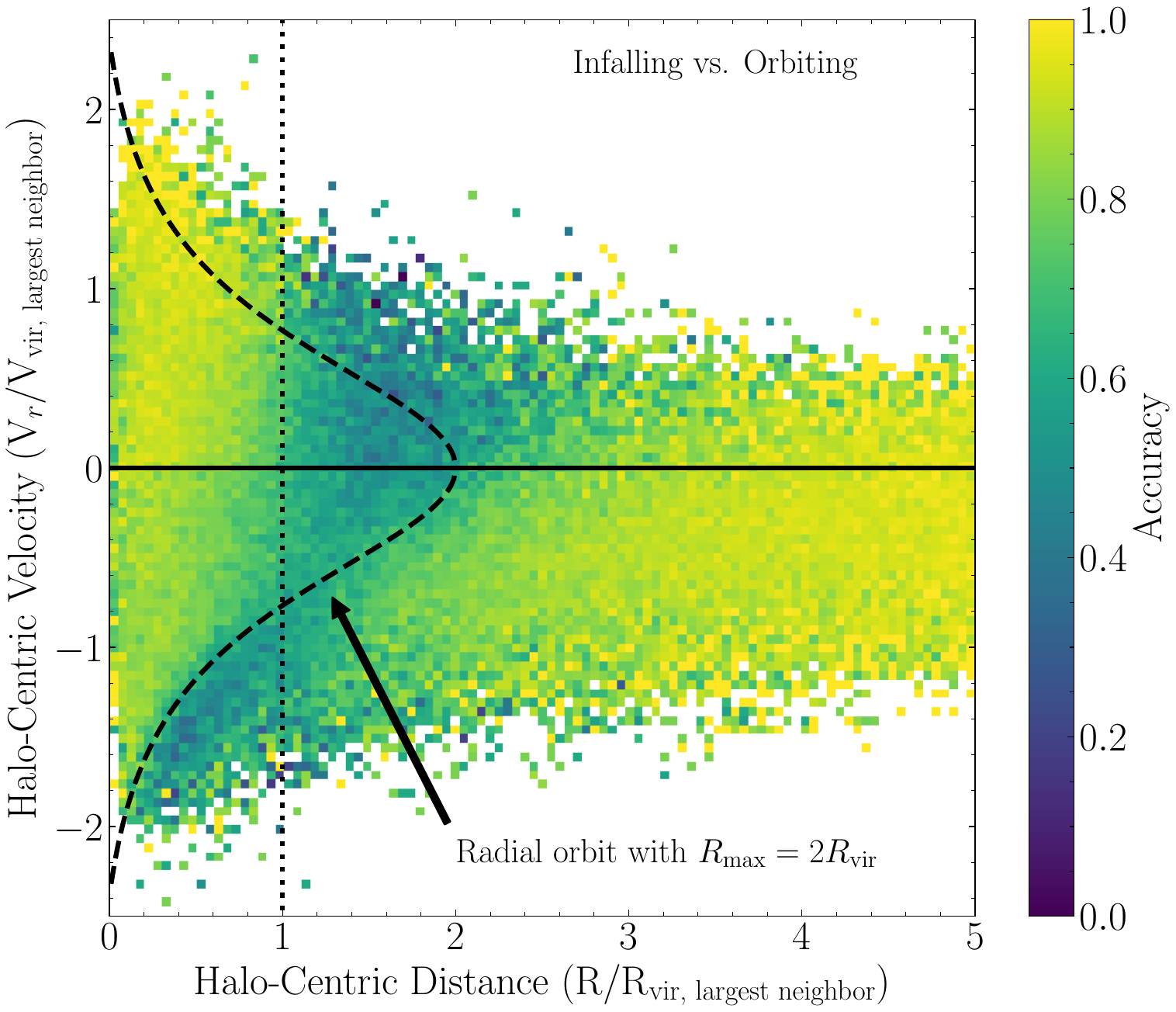}
     \caption{The overall accuracy of the network in recovering infalling versus orbiting status is shown as a function of the relative position and velocity of a halo to its neighbor with the largest tidal influence. The vertical dotted line represents 1 $R_{\text{vir}}$ for the most tidally-influential neighbor. As in Figure \ref{fig:def3a}, the black dotted curve shows the trajectory of a massless particle released at 2 $R_\mathrm{vir}$ falling into a halo.  The lowest accuracy region has shifted from the region surrounding the 1 $R_\mathrm{vir}$ boundary to a region spread across a wide range of physical separations from the most tidally-influential neighbor, roughly aligned with the trajectory shown in the dotted black curve.
     }
     \label{fig:def3_ps}
\end{figure}

With our default network structure (no added weighting by class), we find that orbiting objects are twice as likely to be misclassified than their infalling counterparts, though in making up a smaller fraction of the population, end up accounting for only $\sim 44\%$ of misclassification cases. Infalling objects classified by the network as orbiting are represented by the pink solid-region of Figure \ref{fig:def3_ps}. These objects accounts for 56\% percent of all misclassifications and 100\% of all misclassifications for halo-centric separations greater than $\sim 5$. For both types of misclassification, the greatest number are found in objects lying at or slightly above $R/R_\text{vir, largest} \sim 1$. These misclassifications are most likely attributable to uncertainties in masses and relative positions which make it unclear as to which side of this border region the objects lie on. Without lower uncertainties on mass and 3D separations, there is no information provided to the network capable of effectively distinguishing between the two classes in this region.

\section{Discussion}\label{sec:Discussion}

In this paper, we considered three primary class definitions to break up halos into two populations according to their local environment and dynamical history. Each definition provides us with a different grouping of halos, which may be better suited to different applications. Each definition elucidates different connections between galaxy, halo, and environment, as wells as bringing its own failure modes for classification.

The first classification scheme for current centrals vs. satellites aligns most closely with the standard classification of halos into centrals and satellites. This scheme relies only on information in the current snapshot of the simulation and thus, it is unsurprising that we can accurately recover the categorization given the stellar mass and neighbors information provided to the network in the vast majority of scenarios. In fact, performing a simple search for larger neighbors within a radius around the object (as discussed in sections \ref{subsec:Fixed} and \ref{subsec:CurrentCvS}) also shows a remarkable performance in this classification with an overall accuracy of 86\%. The network provides a greater overall accuracy (90\%), but this small increase showcases that most of the classification information was already captured by the simpler neighbors search.

For example, \cite{Campbell_2015} presented a study of mis-identified centrals and satellites by applying three different group finders to the same mock galaxy catalog, and, in each case, assigning the brightest group member as the central galaxy. As the brightest group galaxy is not always the central, particularly for Milky-Way-sized and larger halos \citep{Bosch_2008,Skibba+2011}, this method is guaranteed to fail in some cases, even when groups are perfectly identified. \cite{Campbell_2015} estimated that this would impact $\sim10\%$ of groups at $10^{13} \Msun$, while errors in the group finding process lead to further misclassifications.

Unlike the process of using a group finder, the method employed in this paper does not guarantee that each group or cluster contains exactly one central. For low mass halos, this can correspond to no objects being singled out as the central. Yet more often, given the default weighting and cut-off between central and satellite, this results in multiple objects classified as centrals within a high mass halo ($M_h > 10^{13} \Msun$). This is especially true in cases when these false centrals are separated by distances of $\gtrsim 0.4$ Mpc from the true central.

Due to this lack of group specification in assignment, the network, when evaluated on a group by group basis for ($M_h \sim 10^{13} \Msun$), recovers the central population to a similar level of completeness as the three group-finder based classifications investigated in \cite{Campbell_2015} ($\sim 85\%$), but with a significantly lower purity ($\sim 70\%$). On the other hand, the neural network approach recovers the satellite population to a much higher purity than the other approach, while maintaining a high completeness. This is likely due to (1) the inclusion of satellites that are not recognized by the group-finder as associated with the group and (2) a tendency for the network to favor satellite purity over central purity in this regime for the default weighting scheme. While one can easily trade off completeness for higher purity (i.e., sacrifice purity in one sample to increase purity in the other), as discussed in Section \ref{subsec:CurrentCvS} and Appendix \ref{sec:AppB}, this result suggests that a group-finder based approach may be more practical for recovering a large and pure sample of group centrals in this mass-regime, than the approach investigated in this paper.

At lower halo masses, particularly for $M_h < 10^{12.5} \Msun$, the cases where the neural-network method assigns multiple objects in a group as centrals becomes a much rarer occurrence. In the $M_h \sim 10^{12} \Msun$ regime, centrals are recovered with a purity of $>95\%$, which is similar to or greater than the purity achieved by any of the group finders investigated in \cite{Campbell_2015} at this regime. Overall, this suggests that additional work is warranted to remove the issue of identifying multiple objects within a group or cluster as the central if the application case involves identifying centrals within group and cluster environments. This might be accomplished by the use of several techniques such as using a graph neural network structure (thereby explicitly linking halos in the network) or through a network that penalizes the assignment of more than one central to a group. In contrast, the neural network as presented in this work is likely sufficient for recovering central and satellite populations in a more generalized context with low overall misclassification rates for both populations.

While many works have sought to recover current central versus satellite status for galaxy-halo pairs, recovering the satellite or dynamical history over a galaxy catalog is a space that is far less explored. The addition of information from previous snapshots into the true classifications presents a challenge for accurately recovering historical satellite status from observable information. In an ideal case, the relative position and velocity of a halo and its neighbors might allow us to reconstruct possible trajectories and thus suggest whether a halo was previously a satellite if it is not so now. Yet, in the case of most surveys, this kind of 3D phase-space information is not available. If we presume access only to the 2D projected distances and redshift separations between halos, we lose much of the constraining power on the individual halos trajectories.

However, our results, as outlined in section \ref{subsec:HistResults}, demonstrate that information about a halo's historical satellite status is contained within the stellar mass and neighbor's information provided to the network, at least when stacked in large quantities.
The network is capable of recovering the true classification from this information for most galaxy-halo pairs across several dex in stellar mass. The greatest performance drop can be seen in Figure \ref{fig:def2_acc} (Bottom), at separations slightly greater that 1 $R_{\text{vir}}$. In these cases, the lack of 3D velocity information makes it especially difficult to distinguish between objects that are moving toward the center of their largest neighbor (have not yet been satellite) and those that were previously within 1 $R_{\text{vir}}$ and are moving away from their neighbor (historical satellites).

The uncertainty introduced by the lack of full velocity information results in a minor ($<2\%$) reduction in overall accuracy between the historical central versus satellite classification scheme and the current central versus satellite scheme. This reduction is primarily the result of failures to classify splashback halos beyond the virial radius as historical satellites. The splashback halo population accounts for $\sim$7\% of halos in the Bolshoi-Planck simulation box. Of these, the default-weighted network for the historical classification scheme recovers $\sim$58\%. This is a substantially higher failure rate for splashback halos than the rest of the halo population.

For applications probing the role that short time-scale interactions with a more-massive halo play on a subhalo and the galaxy within, distinguishing between these splashback halos and centrals with no significant interaction history could have a substantial impact on observed trends. This includes studies of the impact of interaction history on halo or galaxy shapes and sizes, as well as galaxy star formation rates and colors. In these cases, the historical classification scheme may be more relevant than its current counterpart. However, the challenge in distinguishing splashback halos from halos at similar halo-centric distances but with no previous interaction history remains, resulting in a large fraction of interlopers to `true' splashback halos as well as the reverse. Hence, any application of this network to identify splashback halos must take into account the high error-rate of the network on this population. Another consideration is that this scheme makes no distinction based on the length of the interaction between a halo and its more-massive neighbor (i.e., between recent-infalls, objects ejected after a short time, and satellites after many orbits), while the interaction time-scale is known to play a significant role in the processing of both halo and galaxy.

The infalling versus orbiting scheme attempts to add some measure of infall timing to the classification. Yet, recovering the infalling/orbitting status of these subhalos brings additional complications on top of the those seen for the historical central vs. satellite classification. As is shown in Figure \ref{fig:def1}, knowing whether the halo falls to the left or the right side of the dividing line at 1 $R_{\text{vir}}$ when looking at its largest neighbor, is largely sufficient information to determine whether it is current a central or satellite. In contrast, in Figure \ref{fig:def3a}, there is substantial overlap between the orbiting and infalling population in $R/R_{\text{vir, largest neighbor}}$. We find that this overlap does decrease our ability to recover the true classifications for low values of $R/R_{\text{vir, largest neighbor}}$ in contrast to the other classification schemes. Despite this, the network still provides a fairly low misclassification rate for infalling versus orbiting, with a $<4\%$ reduction in overall accuracy compared to the current central versus satellite classification scheme. This suggests that despite the lack of full 3D positions and velocities of individual halos, the network is still able to capture information in the local environment relevant to a halos infalling/orbiting status beyond current satellite status.

The methods explored in this paper are designed for application to a high-completeness low-redshift galaxy survey. Future papers in this series will explore the application of the neural networks trained here to the Galaxy and Mass Assembly (GAMA) catalog.  The GAMA catalog was selected as a local redshift survey with very high completeness as part of its design to investigate galaxy environments \citep{GAMA_main}. However, this method is also suitable for applications to other existing and upcoming spectroscopic galaxy surveys with high completeness (e.g., DESI BGS, WEAVE). In Appendix \ref{sec:AppD} we explore the impact of using a reduced number of neighbors on the network performance, and find that while the performance peaks at 25 neighbors, 5 neighbors is still sufficient for recovering the majority of information regarding current central/satellite status. This is particularly relevant for galaxies near the edges of the survey area, so as to avoid overly limiting the size of the allowed sample by requiring 25 neighbors. 

\section{Conclusions}\label{sec:Conclusions}

Our main conclusions are summarized as follows:
\begin{enumerate}
    \item We present a new method for classifying halos into centrals and satellites (sec \ref{sec:Methods}), which has a baseline error rate of $\sim 10\%$. With small adjustment, the network can be tuned to prioritize different use cases (see appendices). Additional adjustments would be required for direct application to a group or cluster catalog.
    \item We demonstrate that with observable information alone, we can recover the satellite history and orbiting vs. infalling status to an accuracy of $\sim 89\%$ and $86\%$ respectively, providing new insight into halo and galaxy histories and dynamics.
    \item Projection effects are the dominant cause of misclassifications across the three classification cases, with uncertainties in halo mass or a combination of the two factors leading to the remainder of misclassification cases.
\end{enumerate}

\section*{Acknowledgements}

We thank Gurtina Besla, Andrew Hearin, Eduardo Rozo, and Connor Sweeney for their valuable discussions and feedback regarding this manuscript. HB was supported by the Department of Energy HEP-AI program grant DE-SC0023892.  HB and PB were also supported by NASA grant 23-ATP23-0095. The Theoretical Astrophysics Program (TAP) at the University of Arizona provided additional resources to support this work. This research is based upon High Performance Computing (HPC) resources supported by the University of Arizona TRIF, UITS, and Research, Innovation, and Impact (RII) and maintained by the UArizona Research Technologies department. The University of Arizona sits on the original homelands of Indigenous Peoples (including the Tohono O’odham and the Pascua Yaqui) who have stewarded the Land since time immemorial.

The Bolshoi-Planck simulation was performed by Anatoly Klypin within the Bolshoi project of the University of California High-Performance AstroComputing Center (UC-HiPACC; PI Joel Primack). Resources supporting this work were provided by the NASA High-End Computing (HEC) Program through the NASA Advanced Supercomputing (NAS) Division at Ames Research Center. The SMDPL simulation was performed by Gustavo Yepes on the SuperMUC supercomputer at LRZ (Leibniz-Rechenzentrum) using time granted by PRACE, project number 012060963 (PI Stefan Gottloeber).

\section*{Data Availability}

Trained models, as well as the codes used to create them, will be made available online at \url{https://github.com/hbowden-arch/HaloProperties}.
\textsc{UniverseMachine} galaxy catalogs \citep{UM} can be found at \url{https://www.peterbehroozi.com/data.html}.

\bibliographystyle{mnras}
\bibliography{sources}



\appendix

\section{A. Counts in Cylinders Network}\label{sec:AppA}

\begin{figure}[htbp]
     \centering
     \includegraphics[width=0.45\textwidth]{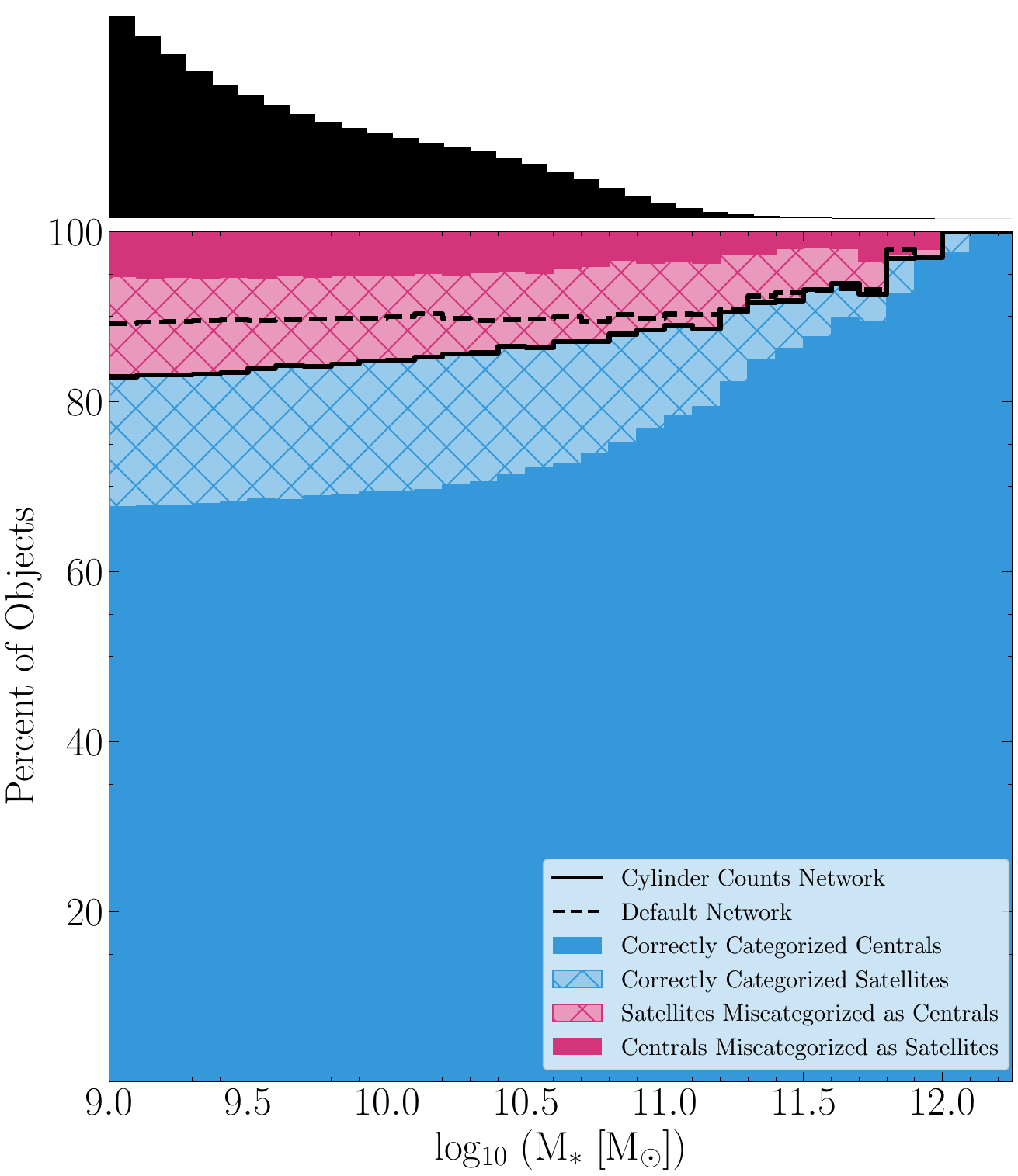}
     \includegraphics[width=0.45\textwidth]{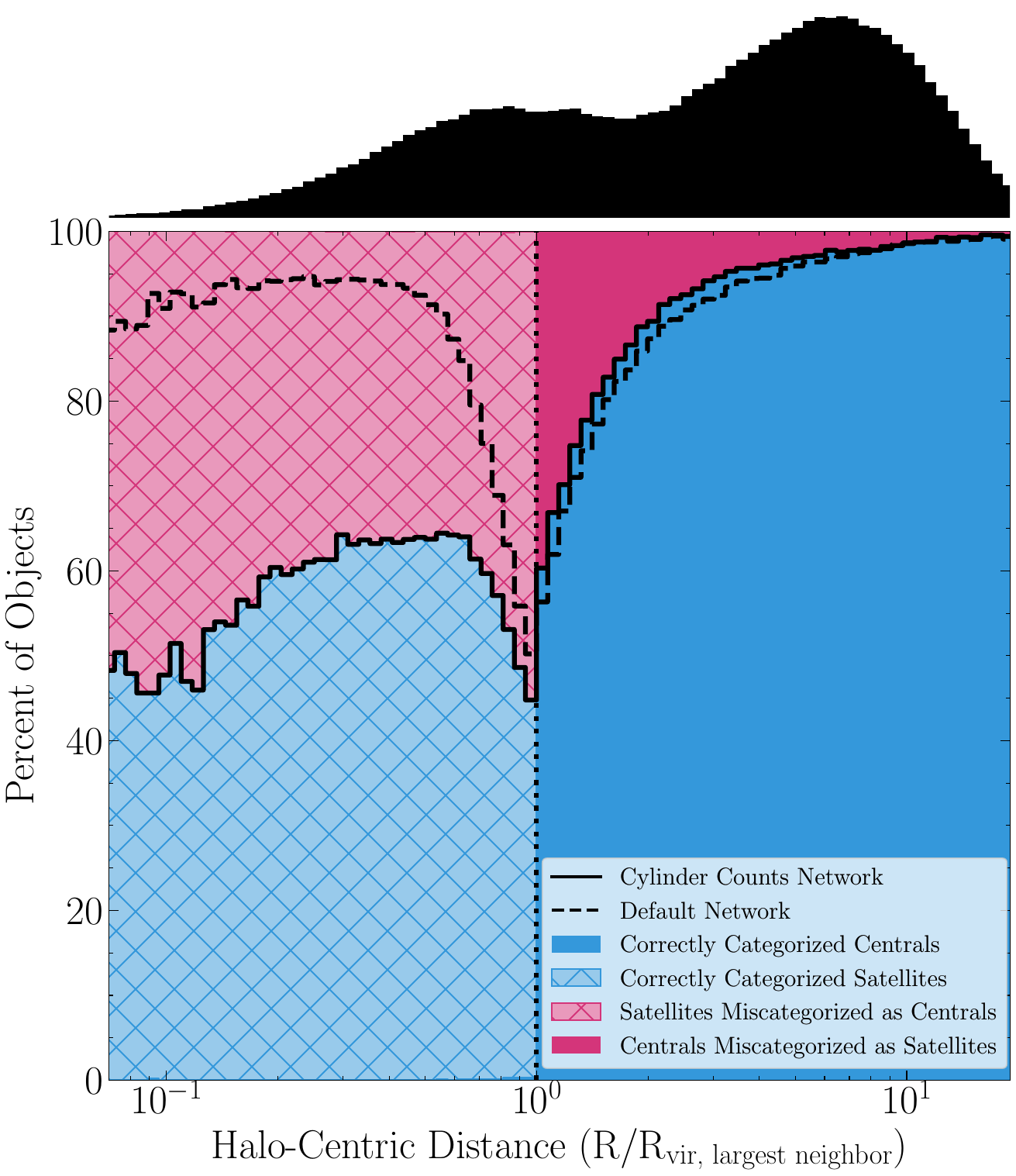}
     \caption{The performance of the cylinder-counts based network on classifying galaxies as centrals of satellites (present) and the number density of halos in the sample (black histograms) is shown as a function of stellar mass (Left) and halo-centric distance (Right). Galaxies are broken into four categories based on both their true and predicted classification as represented by the colored regions, with correctly classified galaxies shown in blue and incorrectly classified in pink. The overall percentage of galaxies that are correctly classified by the network is represented by the black solid line. The dotted vertical line in the bottom plot illustrates the cut between central and satellite as described in Section \ref{sec:Def}. The misclassification rate to the left of this line is much higher than that produced by the $k$NN based network as seen in Figure \ref{fig:def1_acc}.
     }
     \label{fig:cylinder}
\end{figure}

In contrast to the $k$NN measure (Section \ref{subsec:kNN}), counts in cylinders (often known as counts in cells) probes the environment on a fixed distance scale independent of the local density. For the counts in cylinders measure, we defined several circular apertures centered on the target with radii of 0.5 $h^{-1}$Mpc, 1 $h^{-1}$Mpc, 2 $h^{-1}$Mpc, and 5 $h^{-1}$Mpc. This wide range of aperture sizes was selected to provide sensitivity to a wide range of environments. The data are further divided into redshift separation bins of width $|\Delta z| = 250$ km s$^{-1}$, spanning a total range up to $|\Delta z| = 2000$ km s$^{-1}$. This binning strategy allows for improved discrimination against physically unassociated galaxies that are close in projection but lie at larger velocity offsets.

In \citet{Bowden_2023}, we found cylinder counts, combined with the stellar mass of the primary object, to be an effective way of expressing environmental information for estimating halo mass, providing similar accuracy to the $k$NN measure. However, in this work, we find the cylinder counts inputs to generally be inadequate for classifying centrals and satellites. An optimized network provided with cylinder counts inputs achieved an accuracy of only 84.6\% compared to the  89.7\% accuracy achieved by the $k$NN-based network on the same dataset. 

Figure \ref{fig:cylinder}, shows the results of the cylinder-counts network applied to the test dataset. The overall accuracy is much lower than the comparable $k$NN results (see Figure \ref{fig:def1_acc}). In particular, a large proportion of satellites are miscategorized as centrals (pink hashed region). This can be reasonably explained by the lack of mass information attached to individual galaxies for comparison. For example, an object in a high-density environment could be either a satellite or central. Without access to the relative masses of the target and its neighbors, the network cannot accurately distinguish between the two in these environments, and thus defaults to assigning the target as a central as it is the larger of the two populations.

\section{B. Weighting Schemes and Network Confidence}\label{sec:AppB}

\begin{figure}[htbp]
     \centering
     \includegraphics[width=0.45\textwidth]{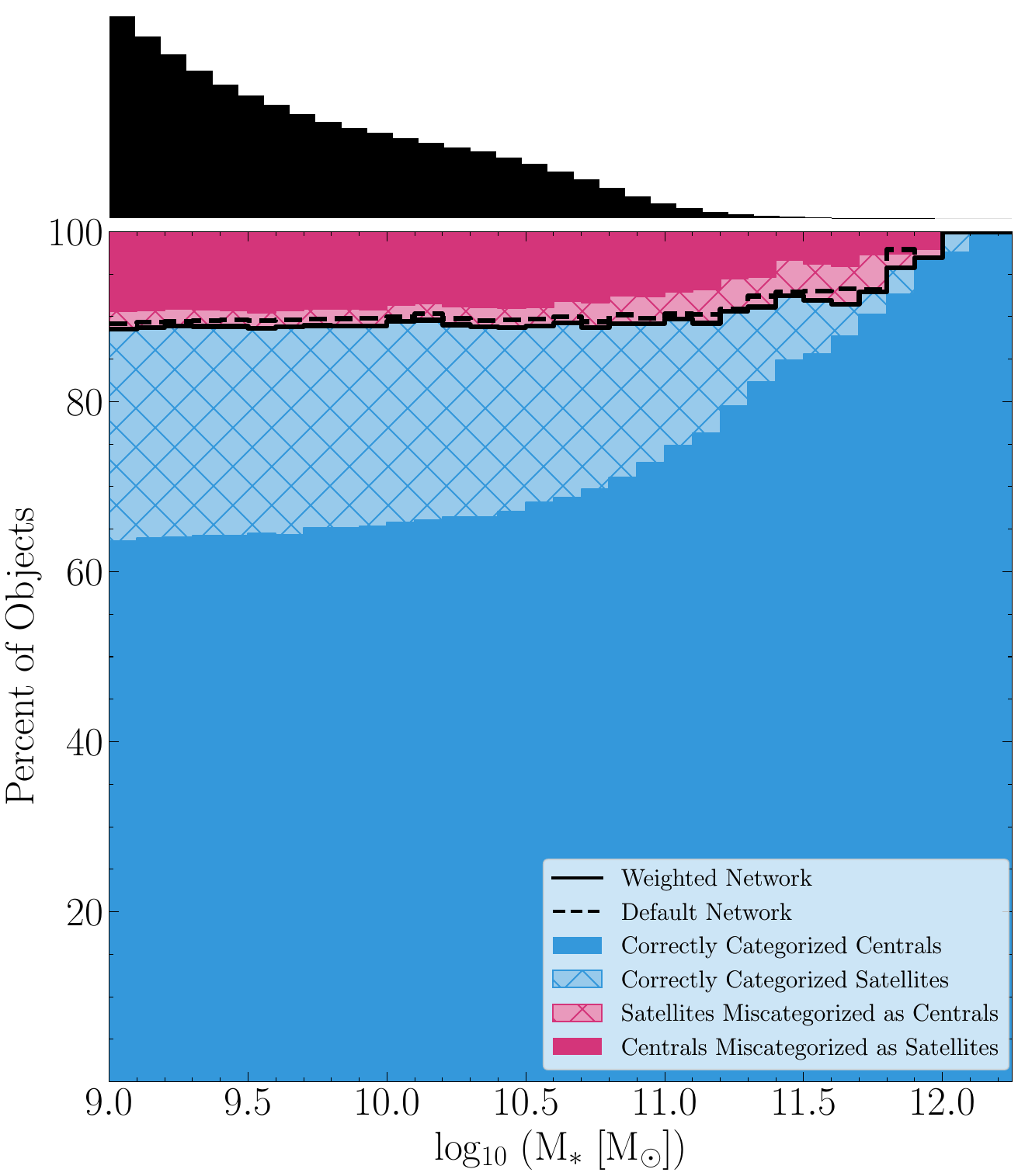}
     \includegraphics[width=0.45\textwidth]{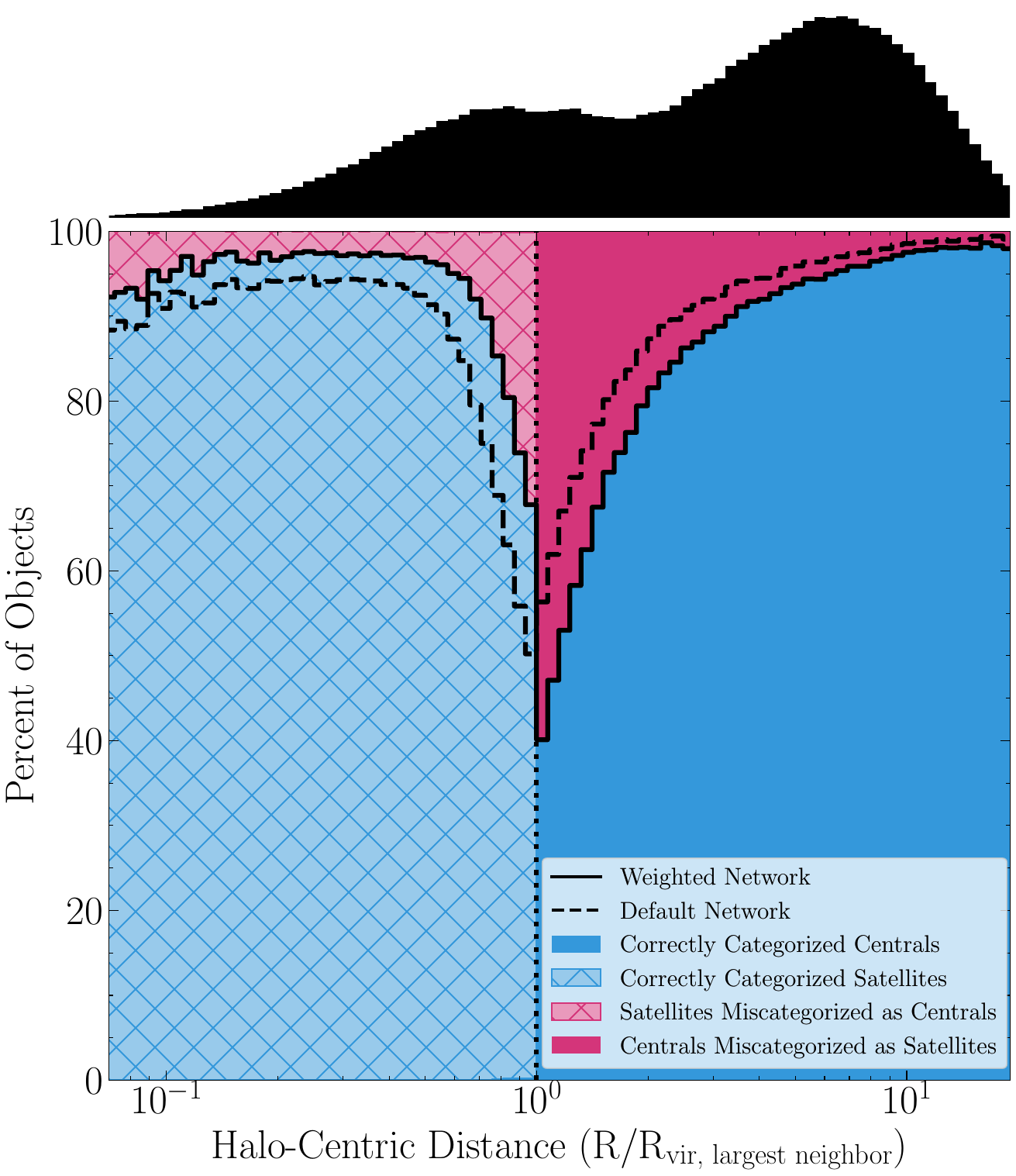}
     \caption{The equivalent of Figure \ref{fig:def1_acc} is shown for the performance of the network when the satellite label is assigned a weight of 2.0 rather than 1.0 during training as a function of stellar mass (Left) and halo-centric distance (Right). Galaxies are broken into four categories based on both their true and predicted classification as represented by the colored regions, with correctly classified galaxies shown in blue and incorrectly classified in pink. The size of the satellites miscategorized as centrals region is reduced compared to Figure \ref{fig:def1_acc}, but with a corresponding increase in the population of centrals miscategorized as satellites. The overall percentage of galaxies that are correctly classified by the network is represented by the black solid line. The dotted vertical line in the right plot shows 1 $R_{\text{vir}}$, or what we consider the halo boundary, for the most tidally-influential neighbor.
     }
     \label{fig:weights_acc}
\end{figure}

Adjusting the weights assigned to the different classes is, in effect, adjusting the prior on which class a target belongs to before any input information is provided to the network. By default, this prior is biased towards assigning central over satellite (and likewise historical central over historical satellite and infalling over orbiting) as centrals make up a larger portion of the training sample by nearly a factor of two. Class weighting allows us to adjust this prior by changing the penalty for misclassification of different classes. For example, applying a class weight of 2 to satellites will result in each incorrectly classified satellite being counted twice in the loss calculation, thus counteracting the initial bias in number of training samples by population. Figure \ref{fig:weights_acc} shows the results of a network trained with this class weighting. Note that the increased penalty for incorrectly classifying satellites as centrals during training has led to a network which is less likely to perform this specific misclassification (pink hatched region), but more likely to incorrectly classify centrals as satellites (pink solid region) when compared to the unweighted case (Figure \ref{fig:def1_acc}).

\begin{figure}[htbp]
    \centering
    \begin{minipage}[t]{0.45\textwidth}
        \centering
        \includegraphics[width=\linewidth]{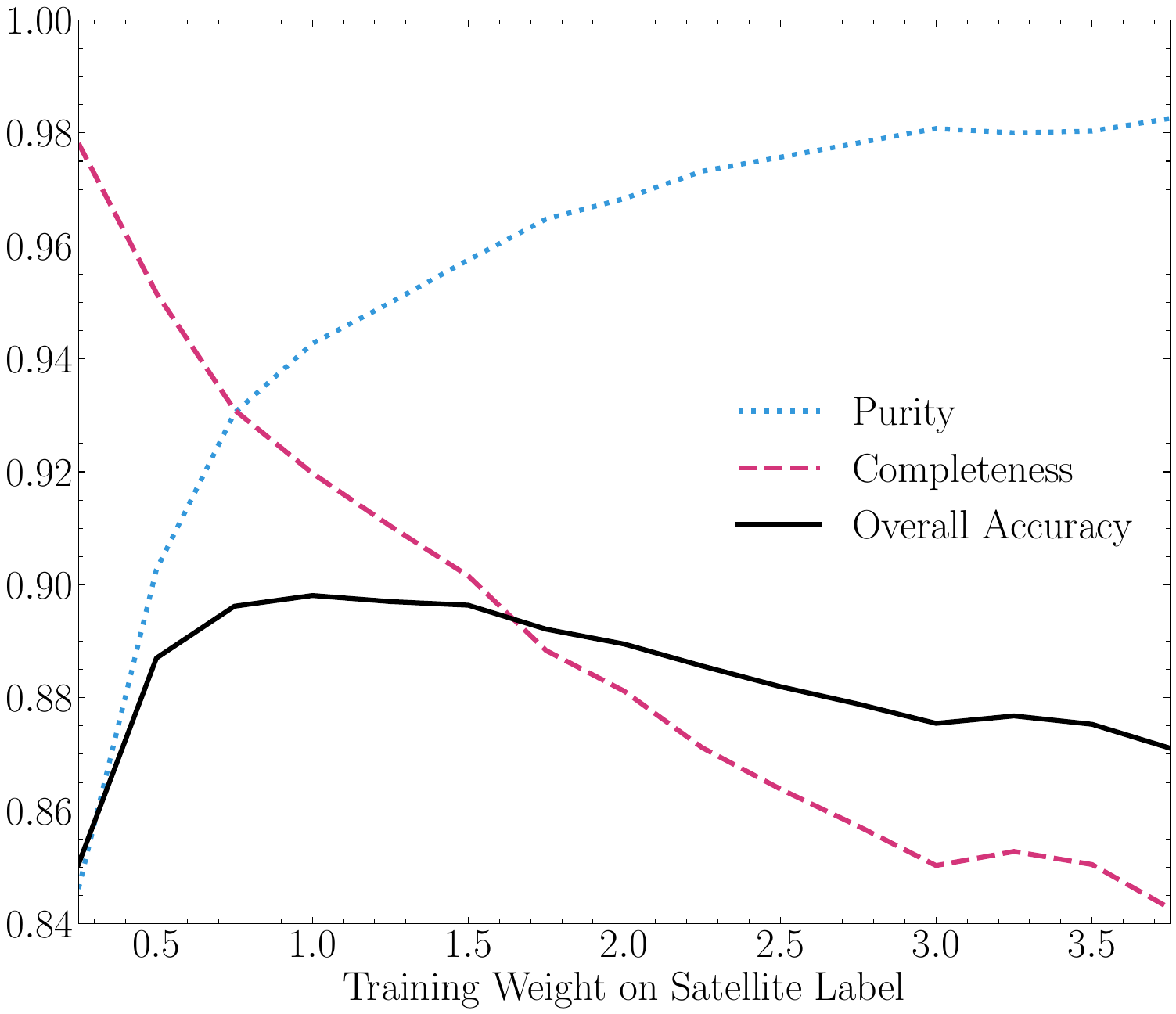}
        \caption{The purity (blue) and completeness (pink) of the recovered central sample are plotted against the weight placed on satellite samples during training, along with the overall accuracy of the network (black) defined as a the fraction of correctly categorized objects when using a cut threshold of 0.5. A larger training weight on the satellite label corresponds generally with an increase in the purity of the central population and a decrease in completeness, while the overall accuracy peaks for weights $\sim 0.75-1.0$.}
        \label{fig:weights}
    \end{minipage}
    \hfill
    \begin{minipage}[t]{0.45\textwidth}
        \centering
        \includegraphics[width=\linewidth]{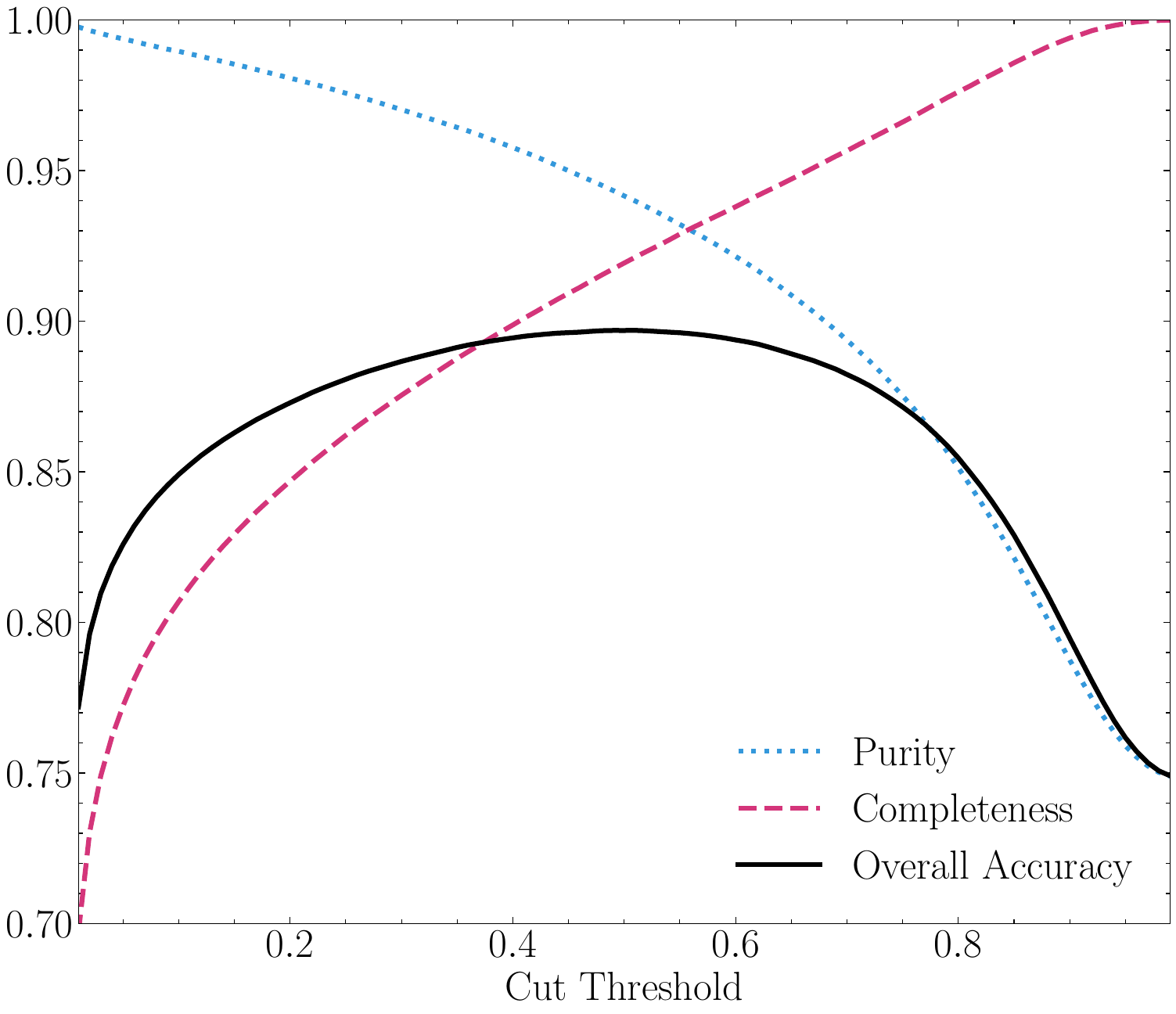}
        \caption{The network output for each object in the current centrals versus satellites is a value between 0 and 1 where the 0 label corresponds to a central and 1 to a satellite. The cut threshold corresponds to network output above which we consider an object to be a satellite. The blue and pink lines here illustrate the resulting purity and completeness of the sample of centrals for a given cut threshold. A cut threshold of 0.5 is used within the main paper.}
        \label{fig:threshold}
    \end{minipage}
\end{figure}

In addition to adjusting an imbalance between classes, class weighting can be used practically as a lever by which to tune the network to prioritize either the purity or completeness of the recovered populations. One should however note that in doing so they are effectively adjusting the prior on their classification. Figure \ref{fig:weights} shows the impact of changing the satellite class weight on the network's performance, while maintaining a central class weight of 1. A training weight of 1 for the satellite class refers to the scenario in which the prior is taken from the population imbalance in the training sample, while a training weight of $\sim 2$ corresponds to an equal weighting of the satellite and central classes. Increasing the training weight on the satellite class increases the purity of the recovered central sample by increasing the penalty associated with misclassifying a satellite as a central. Consequently, the completeness of the central sample decreases as the penalty for misclassifying a satellite as a central outweighs the penalty of misclassifying a central as a satellite. The overall accuracy peaks at $\sim 1$, as the test sample, like the training sample, contains a larger population of centrals than satellites. The historical centrals/satellites and infalling/orbiting cases show highly similar results as they share similar population imbalances in the training and test samples as in the current centrals/satellites case.

Another consideration is the network output, which is not a binary value but rather a continuous value in the range 0 to 1. In the case of classifying current centrals/satellites, this output value represents the likelihood an object is a satellite. Throughout the main paper we assume a cut-off threshold of 0.5, where objects with a likelihood ($\mathcal{L}$) of $<0.5$ are labeled as centrals and those with $\mathcal{L}>=0.5$ as satellites. However, this cut-off threshold is another tunable knob by which we can adjust the purity and completeness of the recovered populations. For example, by requiring object to have $\mathcal{L}<0.25$ to be considered centrals, we select a smaller sample but one which excludes objects that have a higher likelihood of being a satellite. Figure \ref{fig:threshold} shows the trade off between the purity and completeness of the recovered population of centrals as a function of the likelihood upper-limit at which an object is considered a satellite. Note that the purity and completeness of the recovered satellite population will likewise depend on the lower-limit likelihood threshold for assigning the label satellite. It is not necessary that this value is equivalent to the upper-limit for assigning the central, however, if this is not the case, objects falling between the two limits will remain unclassified.

\section{C. 3D-Separation Network Results}\label{sec:AppC}

\begin{figure}[htbp]
     \centering
     \includegraphics[width=0.45\textwidth]{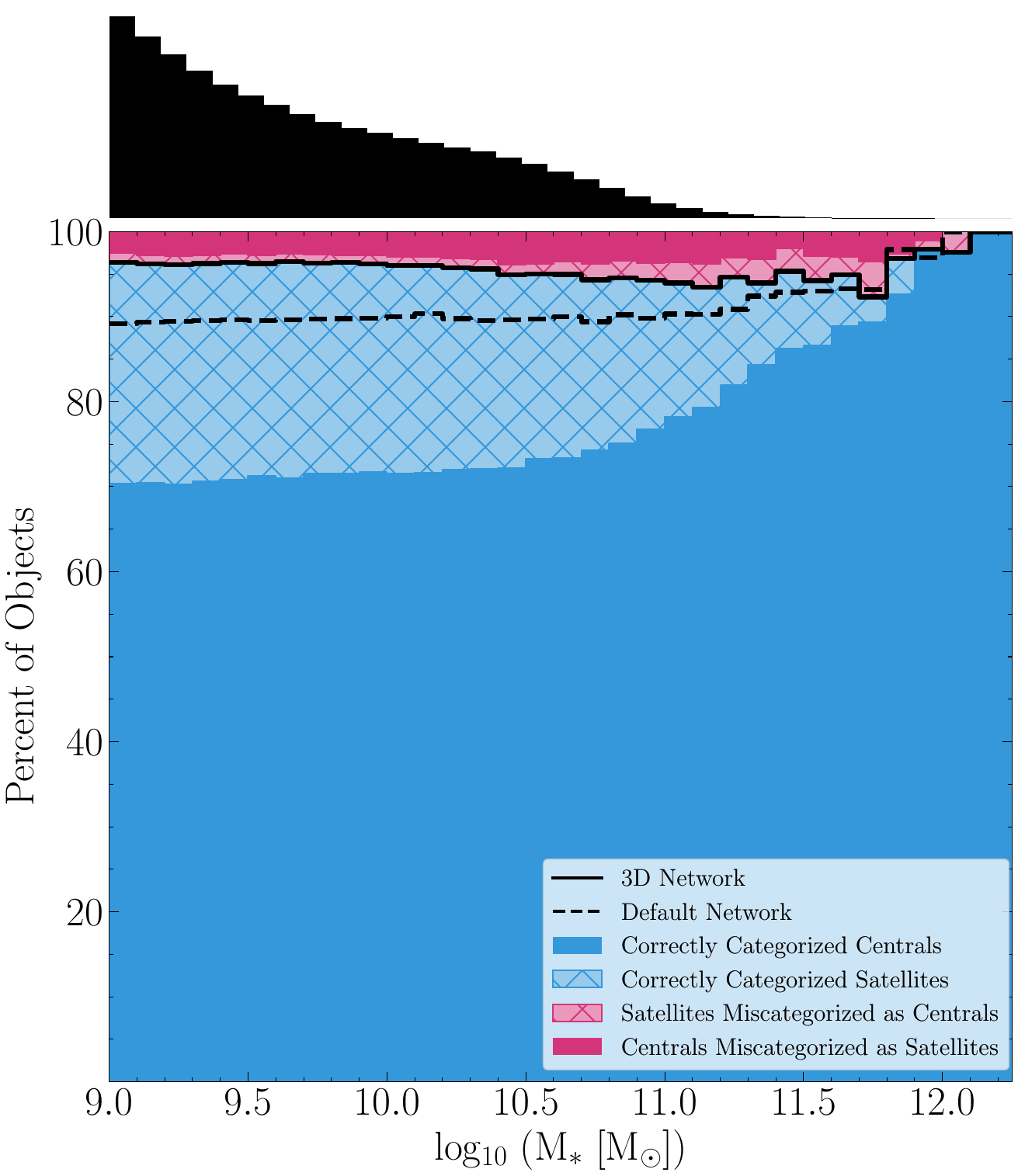}
     \includegraphics[width=0.45\textwidth]{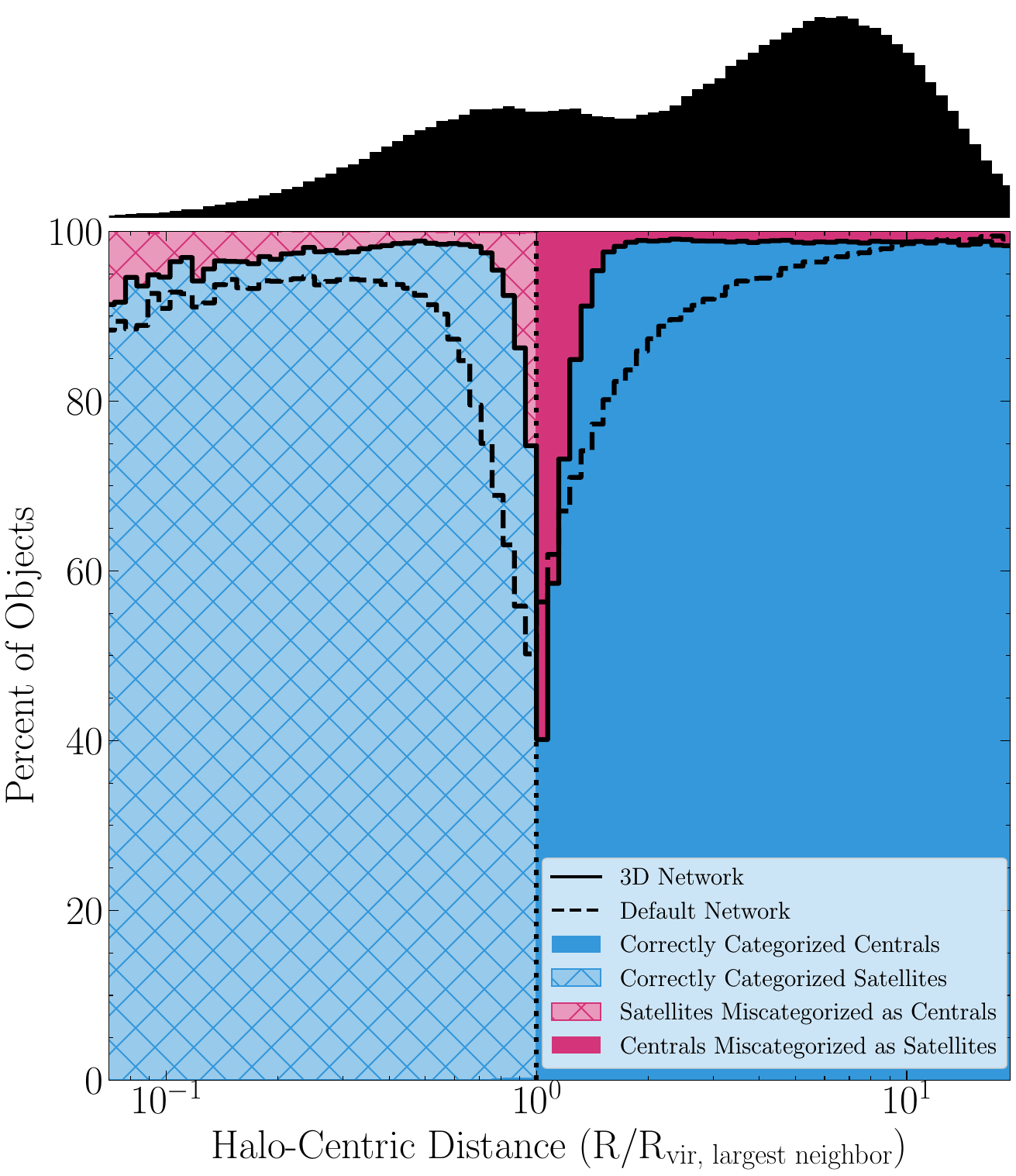}
     \caption{The equivalent of Figure \ref{fig:def1_acc} is shown for the performance of the network when provided with the 3D separation between neighbors rather than the projected separation and redshift-space separation separately. Galaxies are broken into four categories based on both their true and predicted classification as represented by the colored regions, with correctly classified galaxies shown in blue and incorrectly classified in pink. The overall percentage of galaxies that are correctly classified by the network is represented by the black solid line. The misclassification fraction is significantly reduced compared to Figure \ref{fig:def1_acc}, with a sharp decline at 1 $R_{\text{vir}}$ (dotted vertical line), or what we consider the halo boundary, for the most tidally-influential neighbor.
     }
     \label{fig:3D}
\end{figure}

Within the scope of the simulation boxes considered in this paper, we have access to the full 3D positions of our halos. As such, we can test to what extent the use of 2D-projected separations and line-of-sight velocity separations as our metrics for the position of a target relative to its neighbors, such as would be available in an observed catalog, is the limiting factor in recovering the true classification of our targets. In order to do this, we train an additional neural network with the position and line-of-sight information in the input replaced with the 3D physical separation between the objects. Figure \ref{fig:3D} shows the results of the network provided with the full 3D spatial information.

We find a dramatically lower misclassification rate in the 3D case (4.1\%) relative to the network trained on the standard input (10.3\%). In particular, the incidence of satellites miscategorized as centrals decreases to fewer than 5\% of cases, including eliminating nearly all cases at $M_* < 10^{10.5}\Msun$. The remaining cases of misclassification are the result of uncertainty in the halo mass of each object. This primarily effects the classification of objects at $R/R_\text{vir, largest} \sim 1$, as seen in the sharp trough in accuracy around this value in Figure \ref{fig:3D}.

\section{D. Performance with $k < 25$ Neighbors}\label{sec:AppD}

While 25 neighbors provided the best network performance of the values considered (see Section \ref{subsec:Network}), for some surveys probing out to 25 neighbors may prove impractical. Here we explore the accuracy of the $k$NN neural network method with $k<25$ neighbors. For each value of $k$ considered, the network is retrained with inputs corresponding to that number of neighbors. Note that this is distinct from the case of missing neighbors when one might mask certain inputs without retraining the network.

 \begin{figure}
     \centering
     \includegraphics[width=0.45\textwidth]{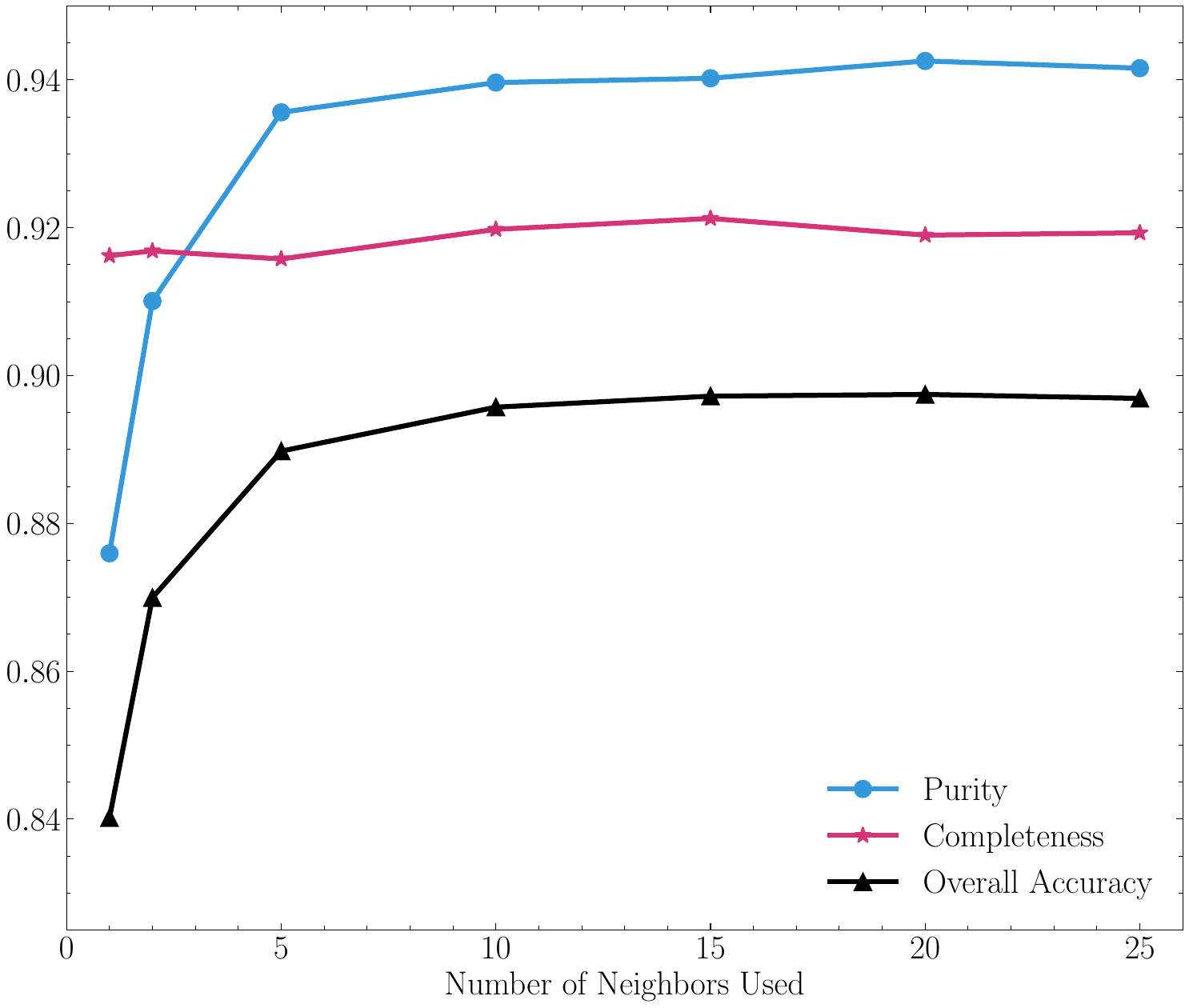}
     \caption{The purity (blue) and completeness (pink) of the recovered central sample are plotted against the number of neighbors provided to network during training and testing, along with the overall accuracy of the network (black) defined as a the fraction of correctly categorized objects when using a cut threshold of 0.5.
     }
     \label{fig:kNN}
\end{figure}

Figure \ref{fig:kNN} shows the performance of 7 networks trained with k=1, 2, 5, 10, 15, 20, and 25 (default) neighbors. The completeness of the recovered central sample (pink stars) is mostly insensitive to the number of neighbors, but the purity of that same sample increases greatly moving from 1 to 2 and 2 to 5 neighbors (blue circles). This leads to a steep improvement in the overall accuracy over the same range. Increasing the value of k above 5 provides a small increase in accuracy, but the majority of the improvement comes in moving from 1 to 5 neighbors. This suggests that the first 5 neighbors provide the relevant information about centrals versus satellites in most cases, and that the information beyond 5 neighbors is not crucial to the network method.


\end{document}